\documentclass[amsmath,amssymb,prd,a4paper,onecolumn,preprint]{revtex4} 
\usepackage[utf8]{inputenc} %         
\usepackage{verbatim}    
\usepackage{graphicx}
\usepackage{dcolumn}
\usepackage{bm}
\usepackage{epsfig}
\usepackage{braket}
\usepackage{epstopdf}
\usepackage{amsfonts}
\usepackage{amssymb,amscd}
\usepackage{color} 
\usepackage[normalem]{ulem} 
\usepackage{longtable} 
\usepackage{multirow}
\newcommand{\be}{\begin{equation}}
\newcommand{\ee}{\end{equation}}
\newcommand{\ben}{\begin{eqnarray}}
\newcommand{\een}{\end{eqnarray}}

\newcommand{\mat}[1]{\mbox{\boldmath{$#1$}}}

\begin{document}

%%%%%%%%%%%%%%%%%%%%%%%%%%%%%%%%%%%%%%%%%%
%%%%%%%%%%%%%%%%%%%%%%%%%%%%%%%%%%%%%%%%%%
\title{The bottom-charmed meson spectrum from a QCD approach based on Tamm–Dancoff Approximation}
%\title{Study of $B_c$ meson spectrum via Coulomb Gauge Approach}
%\title{A  study of the $B_c$ meson spectrum }
%%%%%%%%%%%%%%%%%%%%%%%%%%%%%%%%%%%%%%%%%%
%%%%%%%%%%%%%%%%%%%%%%%%%%%%%%%%%%%%%%%%%%
\author{L.~M.~Abreu}   
\email[]{luciano.abreu@ufba.br}
\affiliation{Instituto de F\'isica, Universidade Federal da Bahia, Campus Universit\'ario de Ondina, Salvador, Bahia, 40170-115, Brazil}

\author{ F.~M.~da~Costa~J\'unior}
\email[]{francisco.miguel@ifsertao-pe.edu.br}
\affiliation{Instituto de F\'isica, Universidade Federal da Bahia, Campus Universit\'ario de Ondina, Salvador, Bahia, 40170-115, Brazil}
\affiliation{Instituto Federal do Sert\~ao Pernambucano, Petrolina, Pernambuco, Brazil}

\author{ A.~G.~Favero}
\email[]{aline.favero@mail.mcgill.ca}
\affiliation{Department of Physics, McGill University, Montr\'eal, QC, H3A 2T8, Canada}

\begin{abstract}

The bottom-charmed meson spectrum is studied in this work via an effective version of the Coulomb gauge QCD Hamiltonian. The Tamm-Dancoff approximation is employed to estimate the energies of the low-lying and radial-excited $B_c$ states with quantum numbers $J^P = 0^{-}, 0^{+}, 1^{-}, 1^{+}, 2^{+}, 2^{-}$. In particular, we analyze the effects of incorporating an effective transverse hyperfine interaction and spin mixing. The Regge trajectories and hyperfine splitting of both $S$- and $P$-wave states are also examined. The numerical results are compared with available experimental data and theoretical predictions of other models. 

\end{abstract}



\maketitle

%%%%%%%%%%%%%%%%%%%%%%%%%%%%%%%%%%%%%%%%%%%%%%%%%%%%%%%%%%%%%%%%%%%%%%%%%%%%%%%%%%%%%
%%%%%%%%%%%%%%%%%%%%%%%%%%%%%%%%%%%%%%%%%%%%%%%%%%%%%%%%%%%%%%%%%%%%%%%%%%%%%%%%%%%%%
\section{Introduction}
\label{Introduction}
%%%%%%%%%%%%%%%%%%%%%%%%%%%%%%%%%%%%%%%%%%%%%%%%%%%%%%%%%%%%%%%%%%%%%%%%%%%%%%%%%%%%%
%%%%%%%%%%%%%%%%%%%%%%%%%%%%%%%%%%%%%%%%%%%%%%%%%%%%%%%%%%%%%%%%%%%%%%%%%%%%%%%%%%%%%

Despite the enormous experimental developments on heavy-hadron physics in recent decades, bottom-charmed $(B_c)$ spectroscopy remains much less known than the charmonium and bottomonium sectors. The reason comes from the fact that since the $B_c$ is a quarkonium bound-state consisting of heavy-quarks with different flavors ($c\overline{b}$ or $b\overline{c}$), the production mechanism demands factories of $c\overline{c}$ and $b\overline{b}$ pairs, which results in a small production rate. 
On the other hand, the different quark-flavor content denies its annihilation into gluons, engendering uniquely weak decays for the pseudoscalar ground-state $B_c(1S)$ and hadronic or radiative transitions for excited states which are below the strong-decay $B D $ threshold. These aspects suggest that $B_c$-states are more stable than their analogs in charmonium and bottomonium families, and therefore are pretty valuable to study heavy-quark dynamics and  understand the dynamics of the strong interaction in a deeper level.

The first observation of $B_c$ meson was performed by CDF Collaboration more than two decades ago~\cite{Abe:1998wi}, with the detection of the pseudoscalar ground-state $B_c(1S)^{+}$. It was confirmed later by other Collaborations~\cite{Aaij:2012dd,Abazov:2008kv}, and is the only state considered as an established particle with recognized quantum numbers, according to Particle Data Group (PDG)~\cite{Tanabashi:2018oca}, with mass $ (6274.9 \pm 0.8)$ MeV.  The other state present in PDG with mass $(6871.0 \pm 1.7)$ MeV and identified as $B_c(2S)^{+}$ has its quantum numbers not confirmed. This is due to the controversy raised by the results from the ATLAS~\cite{Aad:2014laa},   CMS~\cite{Sirunyan:2019osb} and LHCb~\cite{Aaij:2019ldo} Collaborations. ATLAS~\cite{Aad:2014laa} reported the mass $(6842 \pm 4 \pm 5) $ MeV of an observed state consistent with a first radially excited pseudoscalar; while very recently CMS~\cite{Sirunyan:2019osb} and LHCb~\cite{Aaij:2019ldo}  detected two signals consistent with the $2^1S_0$ and $2^3S_1$: for the $2^1S_0$ the LHCb and CMS found the mass being respectively $(6872.1 \pm 1.3 \pm 0.1 \pm 0.8)$ MeV and  $(6871.0 \pm 1.2 \pm 0.8 \pm 0.8)$ MeV. Besides, for the $2^3S_1$, LHCb obtained the mass $(6841 \pm 0.6 \pm 0.1 \pm 0.8) $ MeV, whereas CMS observed the mass difference $m[B_c(2^1S_0)]-m[B_c ^*(2^3S_1)] = (29 \pm 1.5 \pm 0.7)$ MeV. 
Thus, it can be remarked two intriguing features from these reports. The first one is the apparent disagreement between the ATLAS and CMS, LHCb outcomes for the $B_c ^*(2^3S_1)$ meson. One possible explanation is that the peak observed by ATLAS could be the superposition of the $B_c(2^1S_0)$ and $B_c ^*(2^3S_1)$ states, quite narrowly spaced with respect to the resolution of the measurement.  The second one is that the $B_c(2^1S_0)$ state emerges as heavier than the mass $B_c ^*(2^3S_1)$, which is in conflict with theoretical estimations. The plausible justification is that the observed $ B_c ^{\ast}(2S)$ peak has a mass lower than the true value, which remains unknown due to the impossibility of reconstruction of the low-energy photon emitted in the $ B_c ^{\ast +}\rightarrow B_c ^{+} \gamma $~\cite{Sirunyan:2019osb}. Hence, more observations on $B_c $-meson family are expected in the nearest possible future in order to get a detailed characterization of heavy meson spectroscopy.

On theoretical grounds, different perspectives have been consecrated to investigate the $B_c$ meson spectrum as well as to  understand its properties. For example, it can be found  studies in the context of non-relativistic quark models~\cite{Eichten:1994gt,Monteiro:2016ijw,Monteiro:2016rzi,Soni:2017wvy,Akbar:2018hiw,Li:2019tbn,Eichten:2019gig,Chang:2019wpt,Ortega:2020uvc}, relativistic constituent quark models~\cite{Godfrey:1985xj,Zeng:1994vj,Gupta:1995ps,Ebert:2002pp,Ikhdair:2003ry,Godfrey:2004ya}, Quantum Chromodynamics (QCD) sum rules~\cite{Kiselev:1994rc,Wang:2012kw,Aliev:2019wcm}, lattice QCD~\cite{Allison:2004be,Dowdall:2012ab,Mathur:2018epb} and Dyson-Schwinger and the Bethe-Salpeter equations approaches~\cite{Chen:2020ecu}. The point here is that this miscellany of distinct approaches produces a frame to be contrasted with available and future experimental results, which in the end makes possible a compelling comprehension of the $B_c$ phenomenology.

That being so, the present study intends to contribute to the discussion and characterization of the $B_c$ meson spectrum, by employing a different formalism with respect to the preceding analyses mentioned in the previous paragraph. The framework to be utilized is also known as Coulomb gauge QCD model~\cite{Szczepaniak:1995cw,Cotanch:1998ph,LlanesEstrada:1999uh,Szczepaniak:2001rg,LlanesEstrada:2001kr,Ligterink:2003hd,LlanesEstrada:2004wr,LlanesEstrada:2005jf,Szczepaniak:2005xi,General:2007bk,Guo:2007sm,Guo:2008yz,TorresRincon:2010fu,Xie:2013uha,Guo:2014zva,Amor-Quiroz:2017jhs,Abreu:2019adi,Abreu:2020wio}. This formulation is based on the exact QCD Hamiltonian in the Coulomb gauge, which is replaced by an effective Hamiltonian where the original non-perturbative confining and hyperfine interactions can be rearranged into calculable effective potentials between color densities as well as currents. The current quark and gluon field operators are dressed via Bogoliubov-Valatin method. This provides the possibility of using relativistic field theory and many-body techniques such as Tamm-Dancoff and Random Phase approximations. The vacuum is represented as a coherent BCS ground state with quark and gluon Cooper pairs (condensates), and the hadrons interpreted as quasiparticle excitations. This approach has been successfully applied to the description of properties of some types of light and heavy mesons, glueballs, gluelumps, hybrids and tetraquarks~\cite{Szczepaniak:1995cw,Cotanch:1998ph,LlanesEstrada:1999uh,Szczepaniak:2001rg,LlanesEstrada:2001kr,Ligterink:2003hd,LlanesEstrada:2004wr,LlanesEstrada:2005jf,Szczepaniak:2005xi,General:2007bk,Guo:2007sm,Guo:2008yz,TorresRincon:2010fu,Xie:2013uha,Guo:2014zva,Amor-Quiroz:2017jhs,Abreu:2019adi,Abreu:2020wio}. 
So, these reports demonstrate that this model is efficient in retrieving the essential aspects of QCD with a minimal number of free parameters (current quark masses and dynamical constants) and yielding reasonable predictions.

Here we extend the range of applications of the Coulomb gauge QCD model by studying the basic features of $B_c$ mesons within an unified scheme. The interactions between quarks and antiquarks will be treated through an improved confining potential and a transverse hyperfine interaction, whose kernel is a Yukawa--type potential. Estimations for the energies of the low-lying and radial-excited $B_c$ states with quantum numbers $J^P = 0^{-}, 0^{+}, 1^{-}, 1^{+}, 2^{+}, 2^{-}$ are obtained. Also, the Regge trajectories are constructed, and a discussion about the hyperfine splittings of the $S$- and $P$-wave spectroscopy is done. The comparison of our results with other works is performed as well.

The paper is organized as follows. In Section~\ref{Formalism}, we present the Coulomb gauge QCD model within Tamm-Dancoff approximation. Section~\ref{Results} is devoted to show and analyze the numerical calculations of the bottom-charmed meson spectrum, the Regge trajectories and the hyperfine splittings. Concluding remarks are in Section~\ref{Conclusions}. In Appendix we present explicitly the $B_c$ meson spin-orbital wave functions and kernels of the TDA equation of motion used.

%%%%%%%%%%%%%%%%%%%%%%%%%%%%%%%%%%%%%%%%%%%%%%%%%%%%%%%%%%%%%%%%%%%%%%%%%%%%%%%%%%%%%%%
%%%%%%%%%%%%%%%%%%%%%%%%%%%%%%%%%%%%%%%%%%%%%%%%%%%%%%%%%%%%%%%%%%%%%%%%%%%%%%%%%%%%%%%
\section{The Model}
\label{Formalism}
%%%%%%%%%%%%%%%%%%%%%%%%%%%%%%%%%%%%%%%%%%%%%%%%%%%%%%%%%%%%%%%%%%%%%%%%%%%%%%%%%%%%%%
%%%%%%%%%%%%%%%%%%%%%%%%%%%%%%%%%%%%%%%%%%%%%%%%%%%%%%%%%%%%%%%%%%%%%%%%%%%%%%%%%%%%%%%

%%%%%%%%%%%%%%%%%%%%%%%%%%%%%%%%%%%%%%%%%%%%%%%%%%%%%%%%%%%%%%%%%%%%%%%%%%%%%%%%%%%%%
%%%%%%%%%%%%%%%%%%%%%%%%%%%%%%%%%%%%%%%%%%%%%%%%%%%%%%%%%%%%%%%%%%%%%%%%%%%%%%%%%%%%%
%\subsection{Effective Hamiltonian}
%%%%%%%%%%%%%%%%%%%%%%%%%%%%%%%%%%%%%%%%%%%%%%%%%%%%%%%%%%%%%%%%%%%%%%%%%%%%%%%%%%%%%
%%%%%%%%%%%%%%%%%%%%%%%%%%%%%%%%%%%%%%%%%%%%%%%%%%%%%%%%%%%%%%%%%%%%%%%%%%%%%%%%%%%%%

Let us start by introducing the formalism to be used in the analysis of the $B_c$ meson spectrum. It is a Coulomb gauge QCD-inspired model, whose effective Hamiltonian is given by~\cite{LlanesEstrada:1999uh,Szczepaniak:2001rg,LlanesEstrada:2001kr,Ligterink:2003hd,LlanesEstrada:2004wr,LlanesEstrada:2005jf,Szczepaniak:2005xi,Guo:2008yz,TorresRincon:2010fu,Abreu:2019adi,Abreu:2020wio},
\begin{eqnarray}
  H_{eff} & = &  \int d\mathbf{x} \, \Psi^{\dagger}\left(\mathbf{x}\right) \left[-i \mat{\alpha} \cdot \mat{\nabla} + \beta m\right] \Psi\left(\mathbf{x}\right)\nonumber \\
  & &  + H_{C} + H_{T}, 
\label{H_QCD1}
\end{eqnarray}	
where $\Psi$ and $m$ are the current quark field and mass, respectively. The terms $H_{C}$ and $H_{T}$ are the effective couplings associated to the Coulomb and quark hyperfine interactions, i.e.
\begin{eqnarray}
	H_{C} & = &  -\frac{1}{2} \int d\mathbf{x} d\mathbf{y} \rho^{a}\left(\mathbf{x}\right) \hat{V}\left(\vert\mathbf{x} - \mathbf{y}\vert\right) \rho^{a}\left(\mathbf{y}\right),	\nonumber \\
    H_{T} & = & \frac{1}{2} \int d\mathbf{x} \:d\mathbf{y} J_{i}^{a}\left(\textbf{x}\right) \hat{U}_{ij}\left(\mathbf{x}, \mathbf{y}\right) J^{a}_{j}(\textbf{y}),
	\label{H_QCD2}
\end{eqnarray}
where $\rho^{a}(\textbf{x}) =  \Psi^{\dagger}\left(\mathbf{x}\right) T^{a} \Psi\left(\mathbf{x}\right)$ are the color densities and $\mathbf{J}^{a} \left(\mathbf{x}\right)  = \Psi^{\dagger}\left(\mathbf{x}\right) \mat{\alpha} T^{a} \Psi\left(\mathbf{x}\right) $ the quark color currents, 
with $T^{a} $ ($a=1,2,\ldots,8$) being the $SU_{c}(3)$ generators. In the equations above the flavor indices are not explicitly displayed to simplify the notation. Also, it should be mentioned that pure gluonic contributions have been excluded due to the fact that our interest is devoted to the $q\bar{q}$ states. 

We write down below the kernels of the effective couplings in Eq.~(\ref{H_QCD2}) used in the calculations. For the Coulomb longitudinal interaction $H_C$, the kernel is assumed to be an improved confining potential based on Yang-Mills dynamics, which in momentum space is represented as~\cite{LlanesEstrada:2004wr},
\begin{eqnarray}
	V \left( p \right) = \begin{cases} 
	\left(-12.25 \dfrac{ m_g^{1.93}}{p^{3.93}} \right), & \, p < m_g, \\ 
-\dfrac{8.07}{p^2} \dfrac{\ln{\left( \dfrac{p^2}{m_g ^2} + 0.82 \right)^{-0.62}}}{\ln{\left( \dfrac{p^2}{m_g ^2} + 1.41 \right)^{0.8}}}, & \,  p > m_g,  \end{cases}
	  	\label{Coul_Pot2}
\end{eqnarray}
where $m_g$ is a parameter. Although we are not directly dealing with dynamical gluons in our model, we interpret them as responsible for $V \left( p \right)$, obtained from a self-consistent method of the nonabelian degrees of freedom in the presence of static quarks, as noticed by the authors of Ref.~\cite{Szczepaniak:2001rg}. Viewed in this way,  $m_g$ can be interpreted as a dynamical mass scale for the constituent gluons.

 Turning to the term $H_{T}$, it is associated to the quark hyperfine interaction of type $\vec{\alpha}\cdot\vec{\alpha}$ from the second-order coupling between quarks and transverse gluons after integrating out gluonic degrees of freedom. In this sense, the effective transverse hyperfine potential carries the kernel $\hat{U}_{ij}$ which keeps the structure of transverse gauge condition, 
\begin{eqnarray}
\hat{U}_{ij} \left(\mathbf{x}, \mathbf{y}\right) = \left(\delta_{ij} - \frac{\nabla_{i} \nabla_{j}}{\mat{\nabla}^{2}}\right)_{\mathbf{x}} \hat{U}\left(\vert \mathbf{x} - \mathbf{y} \vert\right),
\label{U_ij}
\end{eqnarray}
with $\hat{U}$ being chosen to mimic one-gluon exchange potential. Following the analysis done in Ref.~\cite{LlanesEstrada:2004wr}, in which a Yukawa-type potential appears as the preferred one for reasonable meson descriptions, we choose 
\begin{eqnarray} 
	U \left( p \right) = C_h\begin{cases} 
	(- 24.57) \dfrac{1}{p^2 + m_g ^2}, & p < m_g, \\ 
- \dfrac{8.07}{p^2} \dfrac{\ln{\left( \dfrac{p^2}{m_g ^2} + 0.82 \right)^{-0.62}}}{\ln{\left( \dfrac{p^2}{m_g ^2} + 1.41 \right)^{0.8}}}, &  p > m_g,  \end{cases}
	  	\label{Yuk_pot}
\end{eqnarray}
with the constant $C_h$ standing for the global strength, and the factor $(-24.57)$ being determined by matching the high and low momentum ranges at the scale $m_g$.

Next, we apply an appropriate quark basis in which calculations for meson states are most conveniently made. Following the standard Bogoliubov-Valatin method (see for example Ref.~\cite{LlanesEstrada:2001kr}), we perform the Bogoliubov transformation from the current quark basis to a improved quasiparticle quark basis represented by quasiparticle and antiquasiparticle $B_{\lambda c} ^{(\dagger)} \left(\mathbf{k}\right) , D_{\lambda c} ^{(\dagger)} \left(\mathbf{k}\right)$ operators, which allows us to write the quark field as
\begin{eqnarray}
	\Psi(\mathbf{x}) = \sum_{\lambda \, i}  \int \frac{d^{3}k}{\left(2 \pi\right)^{3}} \left[\mathcal{U}_{\lambda}\left(\mathbf{k}\right) B_{\lambda i}\left(\mathbf{k}\right) + \mathcal{V}_{\lambda}\left(- \mathbf{k}\right) D^{\dagger}_{\lambda i}\left(- \mathbf{k}\right)\right] e^{i \mathbf{k} \cdot \mathbf{x}} \hat{\mat{\epsilon}}_{i},\label{eq:pbcs}
\end{eqnarray}
where $\lambda$ and $i$ denote the helicity and color indices $(i=1,2,3)$, respectively; $ \{ \hat{\mat{\epsilon}}_{c} \}$ is the color vector basis; $\mathcal{U}$ and $\mathcal{V}$ are Dirac spinors forming a complete basis, 
\begin{eqnarray}
	\mathcal{U}_{\lambda}\left(\mathbf{k}\right) &= \frac{1}{\sqrt{2}} \left(\begin{array}{c}
		\sqrt{1 + \sin \phi\left(k\right)} \:\chi_{\lambda}\\
		\sqrt{1 - \sin \phi\left(k\right)} \:\mat{\sigma} \cdot \hat{\mathbf{k}} \:\chi_{\lambda}\end{array}\right),\\
	\mathcal{V}_{\lambda}\left(- \mathbf{k}\right) &= \frac{1}{\sqrt{2}} \left(\begin{array}{c}
		- \sqrt{1 - \sin \phi\left(k\right)} \:\mat{\sigma} \cdot \hat{\mathbf{k}} \: i \sigma_2 \:\chi_{\lambda}\\
		\sqrt{1 + \sin \phi\left(k\right)}  \: i \sigma_2 \:\chi_{\lambda}\end{array}\right),\label{eq:espd}
\end{eqnarray}
with $\chi_{\lambda}$ being the Pauli spinors. 

The Bogoliubov angle $\phi (\arrowvert\mathbf{k}\arrowvert) \equiv \phi_k$ connecting the current and quasiparticle quark bases is obtained by the variational minimization of the quasiparticle vacuum energy $\delta \langle \Omega \vert H \vert \Omega \rangle = 0$, yielding the gap equation 
\ben
k s_k - m c_k  & = & \int_{0} ^{\infty} \frac{q^2}{6 \pi ^2} \left[ s_k c_q  \left(V_1 + 2 W_0 \right) - s_q  c_k  \left(V_0 + U_0 \right)\right]  , 
\label{gap_eq}
\een
where the functions $s_k  \equiv \sin{\phi_k } $ and $c_k \equiv \cos{\phi_k }$ are related to the running quark mass $M (k)$ through the relationship $M (k) = k \tan{\phi_k } $. 
We identify $M (k) \rightarrow m $ at high $k$, while at low $k$ the constituent quark mass is extracted, $M (0) \rightarrow \mathcal{M} $.
The functions $V_0, V_1 $ and $U_0$ denote angular integrals of longitudinal and transverse potentials in the form
\be
F_n (k,q) \equiv \int_{-1} ^{1} dx \; x^n \; F(|\mathbf{k} - \mathbf{q}|), 
\label{ang_int}
\ee
 with $x = \hat{k}\cdot \hat{q}$; and the $W$-function is defined as
\be 
W(|\mathbf{k} - \mathbf{q}|) \equiv U(|\mathbf{k} - \mathbf{q}|) \frac{x (k^2 + q^2) - k q (1 + x^2) }{|\mathbf{k} - \mathbf{q}|^2}. 
\label{W_func}
\ee
%In the following sections, we will also make use of the auxiliary function $Z$: 
%\be 
%Z(|\mathbf{k} - \mathbf{q}|) \equiv U(|\mathbf{k} - \mathbf{q}|) \frac{ 1 - x^2 }{|\mathbf{k} - \mathbf{q}|^2}. 
%\label{W_func}
%\ee

Also, the expectation value of the effective Hamiltonian with respect to the one-quasiparticle state $ \vert q \rangle \equiv  B_{\alpha c} ^{\dagger} \left(\mathbf{k}\right) \vert \Omega \rangle$ engenders the expression that can be identified as the self-energy of the quasiparticle, 
\ben 
\epsilon _{k }  & = &   \langle q \vert H_{eff} \vert q \rangle \nonumber \\
 & = &   m  s_{k} + k c_{k}  - \int_{0} ^{\infty} \frac{q^2}{6 \pi ^2} \left[ s_{k } s_{q} \left(V_0 + 2 U_0 \right) + c_{k} c_{q} \left(V_1 + W_0 \right)\right].
\label{self_en}
\een

It must be observed that a meson in this framework is supposed to be an excited state consisting of a bound state of the quasiparticle and antiquasiparticle. Then, it is useful to introduce the meson creation operator in the TDA scheme,  which is a bosonization method that has been revealed to be a good approximation for a large number of meson families, excluding only the case of the pions. Accordingly, the quasiparticle--antiquasiparticle operator is given by
\begin{eqnarray}
	Q^{\dagger}_{nJP} = \sum_{\lambda \lambda' } \int\frac{d\mathbf{k}}{\left(2\pi\right)^{3}} \Psi^{(nJP)}_{\lambda \lambda' } \left(\mathbf{k}\right) B^{\dagger}_{\lambda}\left(\mathbf{k}\right) D^{\dagger}_{\lambda'}\left(-\mathbf{k}\right),
\label{meson_op}
\end{eqnarray}
where $\Psi^{(nJP)}_{\alpha \beta}$ means the wavefunction corresponding to an open-flavor meson state with total angular momentum $J$, parity $P$ and radial quantum number $n$ (we have omitted the color and flavor indices).

Now the method of calculating the energy levels of mesonic bound states can be expressed.  The energies are obtained via the TDA equation of motion for an open-flavor meson, defined by
\begin{eqnarray}
  	\langle \Psi^{(nJP)} \vert \left[H_{eff}, Q^{\dagger}_{nJP} \right] \vert \Omega \rangle = \left(E_{nJP} - E_{0}\right) \langle \Psi^{(nJP)} \vert Q^{\dagger}_{nJP} \vert \Omega \rangle. 
	\label{TDA_eq}
\end{eqnarray}
This equation can be recast into a more convenient form, by profiting from the rotational invariance of $ H_{eff}$ and constructing the wavefunctions via multiplication of Pauli $\sigma$ matrices by powers of orbital momentum $\hat{k}^l$ to get partial waves. Concerning this last procedure, we indicate to the reader the Appendix A of Ref.~\cite{Abreu:2019adi}, in which the specific case of axial mesons is discussed. Notwithstanding, for completeness we express in detail the wavefunctions exploited in this work, which can be written as (again omitting the flavor indices): 
\begin{eqnarray}
        \Psi _{\lambda \lambda' } ^{(nJP)}\left( \mathbf{k} \right)  & = & \frac{\delta_{i j }^{\rm (color)}}{\sqrt{3}} \, R^{(nJP)} (k) \, \psi_{\lambda \lambda' } ^{(JP)}\left( \mathbf{k} \right), 
	\label{WF_GEN}
\end{eqnarray}
where  $R^{(nJP)} (k) $ is the radial wavefunction; $\psi _{\lambda \lambda' } ^{(JP)} \left( \mathbf{k} \right) $ carries the angular-momentum dependence, and assumes a distinct form according to the nature of the meson state described by the quantum numbers $L, S, J$, which specify the parity $P=(-1)^{L+1}$ and also the charge conjugation $C=(-1)^{L+S}$, if the quark and antiquark have the opposite flavor (equal mass). These wavefunctions are given explicitly in Appendix~\ref{sec:wave}.
After that, we perform the diagonalization of the effective Hamiltonian in the TDA representation, which is undertaken by the computation of the trace of spinor products coming from commutators in the left-hand side of Eq.~(\ref{TDA_eq}). The final expression for the TDA equation of motion
is 
\begin{eqnarray}
 M_{nJP} \, R^{(nJP)} \left(k\right) =  \left(\epsilon_{k}^{b} + \epsilon_{k}^{c} \right) R^{(nJP)}\left(k\right) +  \int\limits_{0}^{\infty} \frac{q^{2} dq}{12 \pi^{2}} \; K^{(JP)} \left(k, q\right) \, R^{(nJP)} \left(q\right), 
\label{TDA_eq_part_wav}
\end{eqnarray}
where $M_{nJP} \equiv E_{nJP} - E_0$ is the energy of the $B_c $ meson state; $\epsilon_{k}^{b} (\epsilon_{k}^{c}$) is the self-energy of the (anti)quasiparticle associated to the $b (c)$ quark; and $K^{(JP)} \left(k, q\right)$ is the kernel bearing the potential terms, which is dependent on the meson quantum numbers. We should remark that several versions of kernels are accessible in literature, written using different basis as well as distinct interaction terms. Until now, the tensor cases with both longitudinal and transverse potentials, however, are not available (at least to our knowledge). In view of these considerations, the relevant kernels obtained for the mesons described by the wavefunctions given by Eqs.~(\ref{WF_PS})-(\ref{WF_PTM}) are expressed in Appendix~\ref{sec:kernels} (Eqs.~(\ref{K_PS})-(\ref{K_T3})).

As a final comment in this Section, we must note that open--flavor mesons, like the $B_c$ mesons, are not eigenstates of charge conjugation, since they have the quark and antiquark with different flavor. Therefore, the total spin $(S)$ is no longer a good quantum number, and spin-singlet and spin-triplet states with $J=L$ can mix. This is the case of axial ($n^3P_1$ and $n^1P_1$) and pseudotensor ($n^3D_2$ and $n^1D_2$) states reported above. A simple mixing prescription for these $J = L$ states is: 
\begin{eqnarray}
\vert nL_L ^{\prime} \rangle & = & \cos{\theta_{nL}}\vert n^1L_L \rangle + \sin{\theta_{nL}} \vert n^3L_L \rangle , \nonumber \\
\vert nL_L  \rangle & = & -\sin{\theta_{nL}}\vert n^1L_L \rangle + \cos{\theta_{nL}} \vert n^3L_L \rangle , 
\label{mix1}
\end{eqnarray}
where $\theta_{nL}$ is the mixing angle and $  nL_L ^{\prime}, nL_L $ are the physical states.  Supposing that the masses of $b$ and $c$--quarks satisfy the limit $m_b>>m_c$, this leads to the extreme heavy--light expression: $  \theta_{nL} \rightarrow \tan{{}^{-1} \sqrt{L/(L+1)} } $, giving $\theta_{nP} \rightarrow  35.3^{\rm o}, \theta_{nD}\rightarrow  39.2^{\rm o}$ .
Here we adopt the following relation between the masses of $(n^3L_L-n^1L_L)$ and  $(  nL_L ^{\prime} - nL_L) $ pairs~\cite{Blundell:1995au}, 
\ben
M(n L_L ) & = & M(n^1 L_L) \cos^2{\theta _{nL}} + M(n^3 L_L) \sin^2{\theta _{nL}} - [M(n^3L_L) - M(n^1 L_L)] \frac{\sin^2{2 \theta _{nL}}}{2 \cos{2 \theta _{nL}}},  \nonumber \\
M(n L_L ^{\prime}) & = & M(n^1L_L) \sin^2{\theta _{nL}} + M(n^3L_L) \cos^2{\theta _{nL}} + [M(n^3 L_L) - M(n^1 L_L)] \frac{\sin^2{2 \theta _{nL}}}{2 \cos{2 \theta _{nL}}} . \nonumber \\
\label{mixing}
\een

\section{Numerical Results}
\label{Results}
%%%%%%%%%%%%%%%%%%%%%%%%%%%%%%%%%%%%%%%%%%%%%%%%%%%%%%%%%%%%%%%%%%%%%%%%%%%%%%%%%%%%%%
%%%%%%%%%%%%%%%%%%%%%%%%%%%%%%%%%%%%%%%%%%%%%%%%%%%%%%%%%%%%%%%%%%%%%%%%%%%%%%%%%%%%%%%

In this Section are exhibited the results for the spectrum of the $B_c$ mesons, generated with the model sketched out above. Briefly, the strategy consists in solving the gap equation (Eq.~(\ref{gap_eq})) for each flavor, in order to get the $k$-dependent  gap angles $\phi_{k}^{b} $ and $\phi_{k}^{c} $; they supply the values of functions $c_{k(q)}^{b(c)} $ and $s_{k(q)}^{b(c)} $ to generate $M_{nJP}$ that solve numerically the TDA equation of motion in Eq.~(\ref{TDA_eq_part_wav}). It should be emphasized the adoption in the calculations  of kernels with interactions represented by an improved confining potential and a transverse Yukawa-type potential playing the role of the exchange of a constituent gluon. 

In the Coulomb gauge QCD model the input parameters to be fitted to the experimental data are the dynamical mass of the constituent gluon $m_g $, the current quark masses of the $b$ and $c$ quarks,  $m_{b}$ and $m_{c}$, and the magnitude of the transverse potential $C_h$. However, as discussed in the Introduction, data for the $B_c$-meson families are scarce at present, despite recent results from the ATLAS~\cite{Aad:2014laa}, CMS~\cite{Sirunyan:2019osb} and LHCb~\cite{Aaij:2019ldo} Collaborations. According to PDG~\cite{Tanabashi:2018oca}, until now there are two $B_c$ mesons observed: the ground pseudoscalar state is the only one considered as an established particle, with mass $M[B_c(1S)^{+}] \approx( 6274.9 \pm 0.8)$ MeV; the other one with mass $(6871.0 \pm 1.7)$ MeV is consistent with a first radially excited pseudoscalar, but quantum numbers are not confirmed. Nevertheless, it should be also mentioned that ATLAS and LHCb Collaborations reported the observations of peaks at $(6842 \pm 4 \pm 5)$ MeV and $(6841 \pm 0.6 \pm 0.1 \pm 0.8)$ MeV, respectively,  which are consistent with the $B_c ^{\ast}(2 ^{3}S_1)$. 
Remarking that the goal here is to extract the basic picture of the $B_c$ meson spectrum, the values of the parameters $( m_b, m_c, m_g, C_h)$ are adjusted to reproduce approximately these reported states, in particular the confirmed $B_c(1S)^{+}$.

%The integrations in Eqs.~(\ref{gap_eq}) and~(\ref{TDA_eq_part_wav}) have been performed with a cutoff  $\Lambda = 6.0 $ GeV. 

%\begin{figure}[htbp]
%\centering
%%\includegraphics[{width=\textwidth}]{ch-rev.eps} \\
%%\includegraphics[{width=\textwidth}]{mdin_ch.eps}
%%\includegraphics[{width=0.6\textwidth}]{ch-rev.eps} \\
%%\includegraphics[{width=0.6\textwidth}]{mdin_ch.eps}
%\caption{Top panel: Running quark mass $M(k)$ as a function of momentum $k$, for different values of $C_h$. Bottom panel: constituent quark mass $\mathcal{M} = M (0) $ as a function $C_h$. The parameters used are $m_g = 600 $ MeV,  $m_f=1$ MeV. The integrations in Eq.~(\ref{gap_eq}) have been performed with a cutoff  $\Lambda = 6.0 $ GeV. }\label{fig:massgap}
%\end{figure}

%%%For completeness, we begin by showing in Fig.~\ref{fig:massgap} the behavior of running quark mass $M(k)$ (obtained from numerical solution of gap equation~(\ref{gap_eq})) with the parameter $C_h$. The constituent quark masses $\mathcal{M}$ can be extracted from the limit $k \rightarrow 0 $, i.e. $\mathcal{M} \equiv M (0) $, while at high scales the current quark mass $m_f$ is recovered. We notice that the growth of the magnitude of transverse potential modifies the value of gap angle coming from gap equation, yielding greater values of constituent quark mass. 

We start by showing in Table~\ref{TABLE-QUARKS} the values of constituent quark masses $\mathcal{M}_{b,c}$ engendered by the current quark masses $m_{b} = 4000$ MeV, $m_{c} = 950$ MeV used as inputs in this subsection. 
The remaining parameters $C_h$ and $m_g $ are taken with different but near values in order to evaluate their impact on the constituent quark masses $\mathcal{M}_{b,c}$, extracted from the limit  $M _{b,c}(k \rightarrow 0) \equiv \mathcal{M}_{b,c}  $. They are chosen obviously keeping in mind  the range that better matches the physical states. It can be seen that the growth of $C_h$ and $m_g $ yields greater values of $\mathcal{M}_{b,c}$, because of the modification of the gap angles coming from solutions of the gap equation. 
We stress that the values of current and constituent quark masses are smaller than in some quark models, due to the contributions from interaction potentials in gap equation~(\ref{gap_eq}) and in the self-energy $\epsilon _k ^{b, c}$ (Eq.~(\ref{self_en})). For a detailed discussion we refer the reader to Refs.~\cite{LlanesEstrada:2004wr,Abreu:2019adi}. 
On this regard, it deserves to be cited that very recent  $(n_f =2+1+1)-$lattice QCD calculations obtained estimations for the  charm quark mass by about 980-995 MeV~\cite{Lytle:2018evc,Hatton:2020qhk}, which are close to the one we utilize.

%The current quark masses $m_f$ (and respective constituent quark masses) used as inputs in the sequence of this subsection are listed in Table~\ref{TABLE-QUARKS}.
%The constituent quark masses $\mathcal{M}_{b,c}$ engendered by the current quark masses $m_{b} = 4000$ MeV, $m_{c} = 950$ MeV used as inputs in this subsection The remaining parameters are kept fixed at the values that generated the outcomes of Fig.~\ref{fig:massgap}, with the evaluation of their impact on the TDA spectrum postponed to the next subsection. Another remark is that the values of $m_f$ in set I are the same as in Ref.~\cite{Abreu:2019adi}; the sets II and III have been chosen  taking different $q_s, q_c, q_b$ masses to evaluate the behavior of computed results with them, as well as to contrast with other works and available experimental data. 

\begin{center}
\begin{table}[h!]
\caption{The constituent quark masses $\mathcal{M}_{b,c}$ engendered by the current quark masses $m_{b} = 4000$ MeV, $m_{c} = 950$ MeV used as inputs in this subsection. $\mathcal{M}_{b,c}$ are obtained from the gap angles $\phi_{k}^{b} $ and $\phi_{k}^{c} $ that solve the gap equation (Eq.~(\ref{gap_eq})), through the relationship $  \lim\limits_{k \rightarrow 0 } M_{b,c} (k) =  \lim\limits_{k \rightarrow 0 } k \tan{\phi_k ^{b,c} }  \equiv \mathcal{M}_{b,c} $. The column ``Set'' denotes the set of parameters $(m_g, C_h)$ used. All quantities are given in MeV, except the value of $C_h$, which is adimensional. }
\vskip1.5mm
\label{TABLE-QUARKS}
\begin{tabular}{c|c|c}
\hline
\hline
Set  $(m_g, C_h)$ & $\mathcal{M}_c $ & $\mathcal{M}_b $ \\ 
\hline
I    $ (600,0.4) $ & 1208 &  4343    
\\
II  $ (650,0.4) $ & 1222 &  4362   
\\
III  $ (700,0.4) $ & 1236 &  4380
\\
IV   $ (700,0.5) $ & 1288 &  4452 
\\
\hline
 Other  & 1000-1600 & 4600-5100 \\
 estimates~\cite{Tanabashi:2018oca,Lytle:2018evc,Hatton:2020qhk} &  &  \\
\hline
\hline
\end{tabular}
\end{table}
\end{center}

\begin{figure}[htbp]
\centering
\includegraphics[{width=0.6\textwidth}]{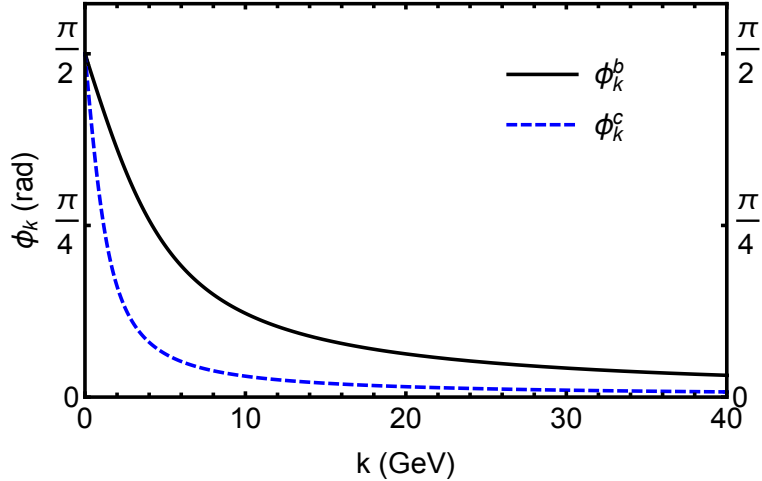} \\
\caption{Bogoliubov angles $\phi_{k}^{b} $ and $\phi_{k}^{c} $, obtained from the solutions of the gap equation (Eq.~(\ref{gap_eq})), as a function of the modulus of the momentum ($k$). We have used the current quark masses $m_{b} = 4000$ MeV, $m_{c} = 950$ MeV and the set III for the parameters $(m_g, C_h)$, in conformity with Table~\ref{TABLE-QUARKS}.}
\label{fig:PhiAngle}
\end{figure}

For the sake of completeness, we briefly discuss the overall momentum-dependence of the Bogoliubov angles for the different flavors obtained from the solutions of the gap equation (Eq.~(\ref{gap_eq})). To this end, in Fig.~\ref{fig:PhiAngle} the solutions $\phi_{k}^{b} $ and $\phi_{k}^{c} $ are plotted as a function of $k$. At higher values of $k$, the solutions exhibit a decreasing exponential behavior, with the $c$-flavor case experiencing a faster lessening. Particularly, the obtention of $\phi_{k}^{b,c} \rightarrow 0 $ in the limit $k \rightarrow \infty$ implies the finiteness of the vacuum energy. At small values of $k$, the solutions present a linear behavior with a negative slope and a sharp peak, yielding $\phi_{k}^{b,c} \rightarrow \pi / 2 $ at $k \rightarrow 0$, which also assures the finite-energy density of the vacuum. Although the specific curves for $\phi_{k}^{b,c}$ obviously depend on the potentials and parameters considered, the point to be stressed is that this formalism yields well-behaved solutions of the gap equation that will be used as inputs in the obtention of the meson spectrum. 

% Since both bottom and charm quarks are heavy, I would suspect
%Bogoliubov angle phi_k is going to be a sharp spike near k=0? I would
%recommend authors to show the plots of phi_k and inserted a few
%comments about it.

%%%%%%%%%%%%%%%%%%%%%%%%%%%%%%%%%%%%%%%%%%%%%%%%%%%%%%%%%%%%%%%%%%%%%%%%%%%%%%%%%%%%%%
%%%%%%%%%%%%%%%%%%%%%%%%%%%%%%%%%%%%%%%%%%%%%%%%%%%%%%%%%%%%%%%%%%%%%%%%%%%%%%%%%%%%%%
\subsection{Mass Spectrum}
%%%%%%%%%%%%%%%%%%%%%%%%%%%%%%%%%%%%%%%%%%%%%%%%%%%%%%%%%%%%%%%%%%%%%%%%%%%%%%%%%%%%%%
%%%%%%%%%%%%%%%%%%%%%%%%%%%%%%%%%%%%%%%%%%%%%%%%%%%%%%%%%%%%%%%%%%%%%%%%%%%%%%%%%%%%%%

Now we report our predictions for the energy levels for the $B_c$ states, extracted from the numerical solutions of the TDA equation in Eq.~(\ref{TDA_eq_part_wav}) considering the different quantum numbers. In Table~\ref{TABLE-MESONS-BC1} are listed the computed masses for ground and radially excited states of $b\bar{c}$ considering the different sets of input parameters of Table~\ref{TABLE-QUARKS}. It gives an overall view of the behavior of computed masses as the parameters $C_h$ and $m_g$ change. In the region of parameter space considered, the augmentation of constituent gluon mass by 100 MeV increases the estimates by about 100-200 MeV, as well as the strengthening of magnitude of transverse potential by 0.1 yields greater masses by about 100-150 MeV.

\begin{longtable}{c|c|c|c|c|c|c}
  \caption{TDA masses of lowest-lying and radially excited $B_c$ states, obtained for $m_c = 950 $ MeV and $m_b = 4000 $ MeV. The column ``Set'' denotes the set of parameters $(m_g, C_h)$ used in conformity with Table~\ref{TABLE-QUARKS}. The masses are given in MeV. Our calculated masses are rounded to 1 MeV.  The mixing angles used are: $\theta_{1P}-\theta_{5P} = 35.3^{\rm o}, \theta_{1D} = 42.5^{\rm o},  \theta_{2D} =  42.2^{\rm o},  \theta_{3D} = 33.2^{\rm o}, \theta_{4D} = 21.1^{\rm o}, \theta_{5D} = 5.2^{\rm o}$. The results that better fit to the observed states are in boldface.}
%\vskip1.5mm
\label{TABLE-MESONS-BC1}
\\
\hline
\hline 
State $(J^P)$ & Set & $E_1$ &  $E_2$ & $E_3$ & $E_4$& $E_5$\\
\hline
 \multirow{4}{*}{$0^-$} & I & 6146 & 6619 & 6986  & 7298 &  7573  \\
& II  & 6212 & 6733 & 7136  & 7477 &  7779  \\
&  {\bf III } & {\bf 6277 }& {\bf 6845 }& 7284  & 7656 &  7983  \\
& IV  & 6417 & 6977 & 7411  & 7779 &  8103   \\
\hline
 \multirow{4}{*}{$0^+$} & I & 6449 & 6852 & 7187  & 7478 &  7740  \\
& II & 6545 & 6989 & 7356  & 7675 &  7961  \\
& III & 6639 & 7123 & 7523  & 7871 &  8181  \\
& IV & 6786 & 7264 & 7659  & 8002 &  8309  \\
\hline
% \multirow{4}{*}{$0^+$} & I & 6411 & 6814 & 7148  & 7439 &  7699  \\
%& II & 6501 & 6945 & 7312  & 7631 &  7916  \\
%& III & 6589 & 7074 & 7474  & 7821 &  8132  \\
%& IV & 6724 & 6724 & 7599  & 7943 &  8250  \\
%\hline
 \multirow{4}{*}{$1^-$} & I & 6154 & 6625 & 6990  & 7302 &  7576  \\
& II &  6222 & 6739 & 7141  & 7482 &  7782  \\
& III &  6288 & {\bf 6853} & 7290  & 7661 &  7988  \\
& IV &  6431 & 6986 & 7418  & 7785 &  8108   \\
\hline
% \multirow{4}{*}{$1^+$ $(S=1)$} & I  & 6434 & 6829 & 7159  & 7448 &  7707  \\
%& II  & 6528 & 6963 & 7325  & 7641 &  7925  \\
%& III  & 6620 & 7095 & 7490  & 7834 &  8142  \\
%& IV  & 6757 & 7225 & 7615  & 7956 &  8260   \\
%\hline
 \multirow{4}{*}{$1 ^+$ } & I  & 6423  & 6821 & 	7153 & 7443 & 7703  \\
& II  & 6516 & 	6955 & 	7318 & 	7635 & 	7920  \\
%& III  & 6605 & 	7085 & 	7482 & 	7827 & 	8136  \\
& III  & 6606 & 7088 &  7488  & 7836 &  8148  \\    
& IV  & 6744 & 	7216 & 	7608 & 	7950 & 	8254   \\
\hline
% \multirow{4}{*}{$1^+$ $(S=0)$} & I   & 6445 & 6837 & 7165  & 7453 &  7711  \\
%& II  & 6540 & 6971 & 7332  & 7647 &  7930  \\
%& III  & 6635 & 7105 & 7498  & 7841 &  8148  \\
%& IV  & 6770 & 7234 & 7622  & 7962 &  8266   \\
%\hline
 \multirow{4}{*}{$1 ^{+ \prime}$ } & I   & 6456 & 6845 & 7171 & 7458 & 7715  \\
& II  & 6552 & 	6979 & 	7339 & 	7653 & 	7935   \\
%& III  & 6650 & 7115 & 	7506 & 	7848 & 	8154  \\
& III & 6.656 & 7121 &  7513  &  7856  & 8164 \\  
& IV  & 6783 & 	7243 & 	7629 & 	7968 & 	8272   \\
\hline
 \multirow{4}{*}{$2^+$ } & I  & 6468 & 6853 & 7178  & 7463 & 7720  \\
& II  & 6568 & 6991 & 7347  & 7659 & 7940  \\
& III  & 6667 & 7127 & 7515  & 7854 & 8159  \\
& IV  & 6805 & 7259 & 7641  & 7976 & 8277   \\
\hline
% \multirow{4}{*}{$2^-$ $(S=1)$} & I  & 6687 & 7034 & 7336  & 7607 & 7853  \\
%& II  & 6806 & 7188 & 7519  & 7815 & 8085  \\
%& III  & 6925 & 7340 & 7701  & 8023 & 8315  \\
%& IV  & 7058 & 7469 & 7824  & 8143 & 8432   \\
%\hline
 \multirow{4}{*}{$2 ^-$} & I  & 6687 & 7023 & 7331 & 7600 & 7845  \\
& II  & 6812 & 7177 & 7512 & 7808 & 8076  \\
& III  & 6931 &  7334 & 7694 &  8015 & 8306 \\
& IV  & 7046 & 7441 & 7815 & 8135 & 8422   \\
\hline
% \multirow{4}{*}{$2 ^-$} & I  & 6687 & 	7028 & 7327 & 7589 & 7829  \\
%& II  & 6809 & 	7182 & 	7507 & 	7797 & 	8058  \\
%& III  & 6928 & 	7337 & 	7689 & 	8002 & 	8288  \\
%& IV  & 7052 & 	7454 & 	7809 & 	8122 & 	8402   \\
%\hline
% \multirow{4}{*}{$2^-$ $(S=0)$} & I  & 6687 & 7032 & 7333  & 7601 & 7845  \\
%& II  & 6807 & 7186 & 7515  & 7809 & 8076  \\
%& III  & 6925 & 7339 & 7697  & 8016 & 8306  \\
%& IV  & 7056 & 7464 & 7819  & 8136 & 8422   \\
%\hline
 \multirow{4}{*}{$2 ^{- \prime}$ } & I  & 6687 & 7043 & 7338 & 7608 & 7853. \\
& II  & 6801 & 7197 & 7522 & 7816 & 8085  \\
& III & 6920 & 7345 & 7704 & 8024 & 8315  \\
& IV  & 7068 & 7492 & 7828 & 8144 & 8432   \\
\hline
% \multirow{4}{*}{$2 ^{- \prime}$ } & I  & 6687  & 7037 & 	7342  & 7619  & 7869 \\
%& II  & 6804  & 	7192  & 	7527  & 	7827  & 	8103  \\
%& III & 6923  & 	7342  & 	7709  & 	8037  & 	8333  \\
%& IV  & 7061  & 	7478  & 	7834  & 	8157  & 	8452   \\
%\hline
\hline
	     \end{longtable}
%\end{table}
%\end{widetext}

On experimental grounds, the set of parameters III seems to generate findings that better fit to the observed states. Although fine tuning of the parameters can give even better outcomes, we believe that set III seems sufficient to generate findings in good conformity with observed states. The spectrum generated for this set is shown schematically in Fig.~\ref{spectrum}.

%They are considerably smaller than the corresponding values ∆1S(cb) = 76 MeV, and ∆2S(cb) = 42 MeV predicted by the quadratic formalism. Moreover, Chen-Kuang [22] predicted ∆1S(cb) = 49.9 MeV, and ∆2S(cb) = 29.4 MeV for their potential with ΛMS = 200 MeV in which the last splitting is almost constant as ΛMS increases. Our predictions for ∆1S(cb) = 68 MeV, and ∆2S(cb) = 35 MeV for the Chen-Kuang potential with ΛMS runs from 100 into 375 MeV . We also find
%∆1S(cb) = 67 MeV, and ∆2S(cb) = 33 MeV for the Igi-Ono potential with ΛMS = 300 MeV and b = 16.3. The present model has the following features: (1) The present potential
%predicts smaller ∆1S and ∆2S than the other potentials do for cb system and the present ∆1S and ∆2S do not depend on ΛMS more sensitively (2) The experimental cb spilitting can be
%repoduced for the preferred range of ΛMS runs from 100 into 375 MeV . Table VII reports our results using SLNET compared to other formalisms.
%
\begin{widetext}
\centering
\begin{figure}[th]
\centering
\includegraphics[{width=0.9\textwidth}]{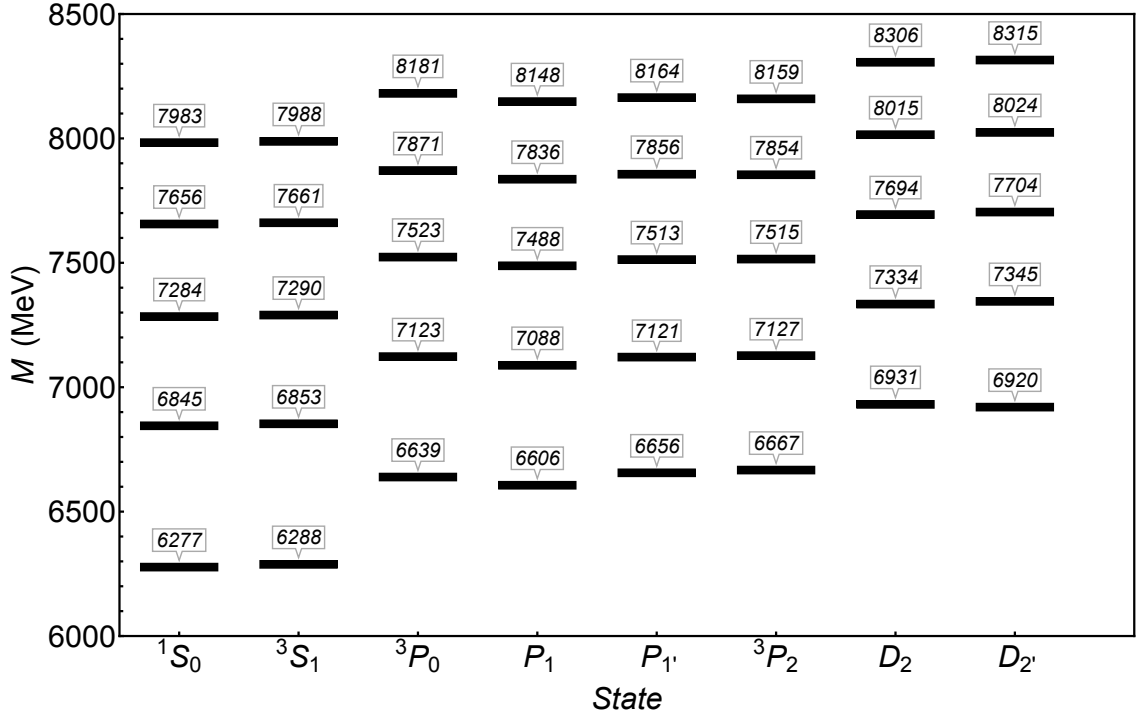} 
\caption{$B_c$ spectrum generated for the set of parameters III.}
\label{spectrum}
\end{figure}
\end{widetext}

It is also noteworthy to evaluate our predictions in light of other works existent in the literature. To this end, in Table~\ref{TABLE-MESONS-BC2} our calculated $B_c$ masses with the set of parameters III are compared with other theoretical results and available experimental data. Stated explicitly, the theoretical frameworks employed in these other studies are: relativized constituent quark model with the presence of a linear confining potential and a color Coulomb interaction~\cite{Godfrey:1985xj}; constituent quark model in heavy quark symmetry limit with scalar confining and vector Coulomb potentials ~\cite{Zeng:1994vj}; non-relativistic quark model (NRQM) consisting of a confinement potential and one gluon exchange potential~\cite{Monteiro:2016ijw}; nonrelativistic linear potential model with a spin-dependent interaction~\cite{Li:2019tbn}.

\begin{widetext}
 
%\begin{table}
%\centering
 \begin{longtable} {c|c|c|c|c|c|c|c|c}
     \caption{Last Column: TDA masses of lowest-lying and radially excited $B_c$ states obtained for the set of parameters III in Table~\ref{TABLE-QUARKS}. Other columns: predictions of other works existent in the literature and available experimental data. The $B_c(1^1 S_0 )$ is the only established particle and its mass has been taken from PDG~\cite{Tanabashi:2018oca}. The states observed by ATLAS~\cite{Aad:2014laa}, CMS~\cite{Sirunyan:2019osb} and LHCb~\cite{Aaij:2019ldo} are not considered as well-established by PDG; quantum numbers of the so-called $2^1 S_0$  need to be confirmed. The masses are given in MeV.} 
\label{TABLE-MESONS-BC2}
\\
\hline
\hline 
State & $ J^P$  & Ref.~\cite{Godfrey:1985xj} & Ref.~\cite{Zeng:1994vj} & Ref.~\cite{Monteiro:2016ijw} & Ref.~\cite{Akbar:2018hiw}  & Ref.~\cite{Li:2019tbn} &  Exp. Data & Our Results \\
	  &        &        &           &     &       &  &   (${}^{\ast}$\cite{Aad:2014laa}; ${}^{\dag}$\cite{Sirunyan:2019osb}; ${}^{\ddag}$\cite{Aaij:2019ldo})   &  (Set III) \\
	  \hline
         $ B_c(1^3 S_1 )$&$1^-$ & 6338 & 6340 & 6357 & 6314 & 6326 &   $\cdots$  & 6288 \\
         $ B_c(1^1 S_0 )$&$0^-$ & 6271 & 6260 & 6275 & 6274 & 6271 & 6275 (PDG) & 6277 \\
     \hline
	     $ B_c(2^3 S_1 )$&$1^-$ & 6887 & 6900 & 8697 & 6855 & 6890 &  $6842^{\dag}; 6841^{\ddag}$ & 6853 \\
	     $ B_c(2^1 S_0 )$&$0^-$ & 6855 & 6850 & 6862 & 6841 & 6871 &  $6842^{\ast}; 6871^{\dag}; 6872^{\ddag}$ & 6845 \\
	  \hline
	     $ B_c(3^3 S_1 )$&$1^-$ & 7272 &7280  & 7333 & 7206 & 7252 &   $\cdots$      & 7290 \\
	     $ B_c(3^1 S_0 )$&$0^-$ & 7250 & 7240 & 7308 & 7197 & 7239 &    $\cdots$     & 7284 \\
	  \hline
%	 $ B_c(4^3 S_1 )$ &$1^-$ &   & 7580 & 7495 & 7550 & 7661 \\
%	 $ B_c(4^1 S_0 )$ &$0^-$ &   & 7550 & 7488 & 7540 & 7656 \\
%	  \hline
%	 $ B_c(5^3 S_1)$ &$1^-$ &   & --       &      & 7813    & 7988 \\
%	 $ B_c(5^1 S_0)$ &$0^-$ &   & --       &     & 7805     & 7983 \\
%	  \hline
	 $ B_c(1^3 P_2) $ &$2^+$ & 6768 & 6760 & 6737 & 6753 & 6787 &   $\cdots$     & 6667 \\
	 $ B_c(1P_1 ^{\prime})$ &$1^+$ & 6750  & 6740 & 6734 &  6744  & 6776 &   $\cdots$     & 6656 \\
	 $ B_c(1P_1 )   $ &$1^+$ & 6741 & 6730 & 6686 & 6725 & 6757  &   $\cdots$     & 6606 \\
	 $ B_c(1^3 P_0 )$ &$0^+$ & 6706 & 6680 & 6638 & 6701 & 6714 &     $\cdots$   & 6639 \\
 \hline
	 $ B_c(2^3 P_2)  $&$2^+$ & 7164 & 7160 & 7175 & 7111 & 7160 &     $\cdots$   & 7127 \\
	 $ B_c(2 P_1 ^{\prime})$&$1^+$ & 7150  & 7150 & 7173 & 7098   & 7150 &  $\cdots$      & 7121 \\
	 $ B_c(2 P_1)    $&$1^+$  & 7145 & 7140 & 7137 & 7105 & 7134 &    $\cdots$    & 7088 \\
	 $ B_c(2^3 P_0)  $&$0^+$ & 7122 & 7100 & 7084 & 7086 & 7107 &      $\cdots$  & 7123 \\
	    \hline
	 $ B_c(3^3 P_2 ) $&$2^+$ &   $\cdots$  & 7480 & 7575 & 7406 & 7464 &     $\cdots$   & 7515 \\
	 $ B_c(3 P_1 ^{\prime})$&$1^+$ &  $\cdots$    & 7470 & 7572 & 7393   & 7458 &   $\cdots$     & 7513 \\
	 $ B_c(3 P_1 )   $&$1^+$ &   $\cdots$   & 7460 & 7546 & 7405  & 7441 &    $\cdots$    & 7488 \\
	 $ B_c(3^3 P_0 ) $&$0^+$ &   $\cdots$  & 7430  & 7492 & 7389& 7420  &    $\cdots$    & 7523 \\
	   \hline
%	 $ B_c(4^3 P_2 ) $&$2^+$ &   & 7760 &     & 7732 & 7854 \\
%	 $ B_c(4 P_1 ^{\prime})$&$1^+$  &   & 7740 &      & 7727 & 7848 \\
%	 $ B_c(4 P_1 )   $&$1^+$ &    & 7740  &     & 7710  & 7827 \\
%	 $ B_c(4^3 P_0 ) $&$0^+$ &   & 7710 &    & 7693   & 7821 \\
%	    \hline
%	 $ B_c(5^3 P_2) $ &$2^+$ &   &          &          &      &      \\
%	 $ B_c(5P_1 ^{\prime})$ &$1^+$ &     &          &          &      & 8154 \\
%	 $ B_c(5P_1 )   $ &$1^+$  &   &          &          &      & 8135  \\
%	 $ B_c(5^3 P_0 )$ &$0^+$  &    &         &          &      &      \\
% \hline
%	 $ B_c(1^3 D_3 ) $&$3^-$   & 7045 &      &  6998   & 7030   &      \\
	 $ B_c(1 D_2 ^{\prime})$&$2^-$   & 7036 &  $\cdots$   & 7003   &  6984   & 7032   &   $\cdots$     & 6920 \\
	 $ B_c(1 D_2 )   $&$2^-$   & 7041 &   $\cdots$     & 6974 &   6986 & 7024     &      $\cdots$  & 6931 \\
%	 $ B_c(1^3 D_1 ) $&$1^-$   & 7028 &      & 6964   & 7020    &      \\
	    \hline
%	 $ B_c(2^3 D_3 ) $&$3^-$ &         &      &  7302   & 7348      &      \\
	 $ B_c(2 D_2 ^{\prime})$&$2^-$ &  $\cdots$   &   $\cdots$     & 7408 & 7293   & 7347  &    $\cdots$    & 7345 \\
	 $ B_c(2 D_2 )   $&$2^-$       &  $\cdots$   &    $\cdots$    & 7385 & 7294    & 7343  &   $\cdots$     & 7334 \\
%     $ B_c(2^3 D_1 ) $&$1^-$       &   &      &  7280  & 7336  &      \\
	      \hline
%	 $ B_c(3^3 D_3 ) $&$3^-$ &   &      & 7570   & 7625   &       \\
	 $ B_c(3 D_2 ^{\prime})$&$2^-$ & $\cdots$   &  $\cdots$   & 7783 & 7562  & 7623  &   $\cdots$     & 7704  \\
	 $ B_c(3 D_2 )   $&$2^-$ &  $\cdots$   &    $\cdots$     & 7781 &  7563   & 7620  &  $\cdots$      & 7694  \\
%	 $ B_c(3^3 D_1 ) $&$1^-$ &   &      & 7553   & 7611   &       \\
%	      \hline
%	 $ B_c(4^3 D_3 ) $&$3^-$ &   &      &      &      &       \\
%	 $ B_c(4 D_2 ^{\prime})$&$2^-$ &   &      &      &      & 8023  \\
%	 $ B_c(4 D_2 )   $&$2^-$ &   &      &      &      & 8016  \\
%	 $ B_c(4^3 D_1 ) $&$1^-$ &   &      &      &      &       \\
%	      \hline
%	 $ B_c(5^3 D_3 ) $&$3^-$ &   &      &      &      &        \\
%	 $ B_c(5 D_2 ^{\prime})$&$2^-$ &   &      &      &      & 8315    \\
%	 $ B_c(5 D_2 )   $&$2^-$ &   &      &      &      & 8306    \\
%	 $ B_c(5^3 D_1 ) $&$1^-$ &   &      &      &      &        \\
\hline
	   \hline	    
 %\end{tabular}
 \end{longtable}
 \end{widetext}

We stress that our findings, specially for $B_c(1S) $ and $ B_c ^{\ast}(2S)$, present a very good fit with the measured data when the experimental errors are bore in mind. But this comparison must be done with care, because the observed $ B_c ^{\ast}(2S)$ peak has a mass lower than the true value, which remains unknown 
due to the impossibility of reconstruction of the low-energy photon emitted in the $ B_c ^{\ast +}\rightarrow B_c ^{+} \gamma $, as pointed out in Ref.~\cite{Sirunyan:2019osb}. 
Moreover, the mass of the first radial excitation $B_c(2S)$ is heavier than the ground state $B_c(1S)$ by about 557 MeV,  which is fairly good in light of experimental observations, keeping the fact that $B_c(2S)$ is not yet well-established according to PDG~\cite{Tanabashi:2018oca}. 

Furthermore, it can be seen that our outcomes get the $B_c$ spectrum in reasonable concordance with other potential model predictions. In general, the masses predicted by us for the low-lying states have a difference  with respect to  previous works ranging from a few MeV up to tens of MeV. The exceptions having larger mass deviations are the $1^3 P_2, 1 P_1, 1 P_1 ^{\prime}  $ states.  For higher mass states, bigger discrepancies among the predictions are evident,  but most of our results are between the lower and upper values reported in Table~\ref{TABLE-MESONS-BC1}. Particularly, our results for $1S, 2S, 3S, 2P-$wave states are up to a few tens of MeV discrepant with those with relativized constituent quark model from Ref.~\cite{Godfrey:1985xj}, while for $1P_2,1P_1 ^{(\prime)}, 1D-$wave states are about $100-140 $ MeV smaller.

%%%%%%%%%%%%%%%%%%%%%%%%%%%%%%%%%%%%%%%%%%%%%%%%%%%%%%%%%%%%%%%%%%%%%%%%%%%%%%%%%%%%%%
%%%%%%%%%%%%%%%%%%%%%%%%%%%%%%%%%%%%%%%%%%%%%%%%%%%%%%%%%%%%%%%%%%%%%%%%%%%%%%%%%%%%%%
\subsection{Regge Trajectories}
%%%%%%%%%%%%%%%%%%%%%%%%%%%%%%%%%%%%%%%%%%%%%%%%%%%%%%%%%%%%%%%%%%%%%%%%%%%%%%%%%%%%%%
%%%%%%%%%%%%%%%%%%%%%%%%%%%%%%%%%%%%%%%%%%%%%%%%%%%%%%%%%%%%%%%%%%%%%%%%%%%%%%%%%%%%%%

\begin{figure}[htbp]
\centering
\includegraphics[{width=0.6\textwidth}]{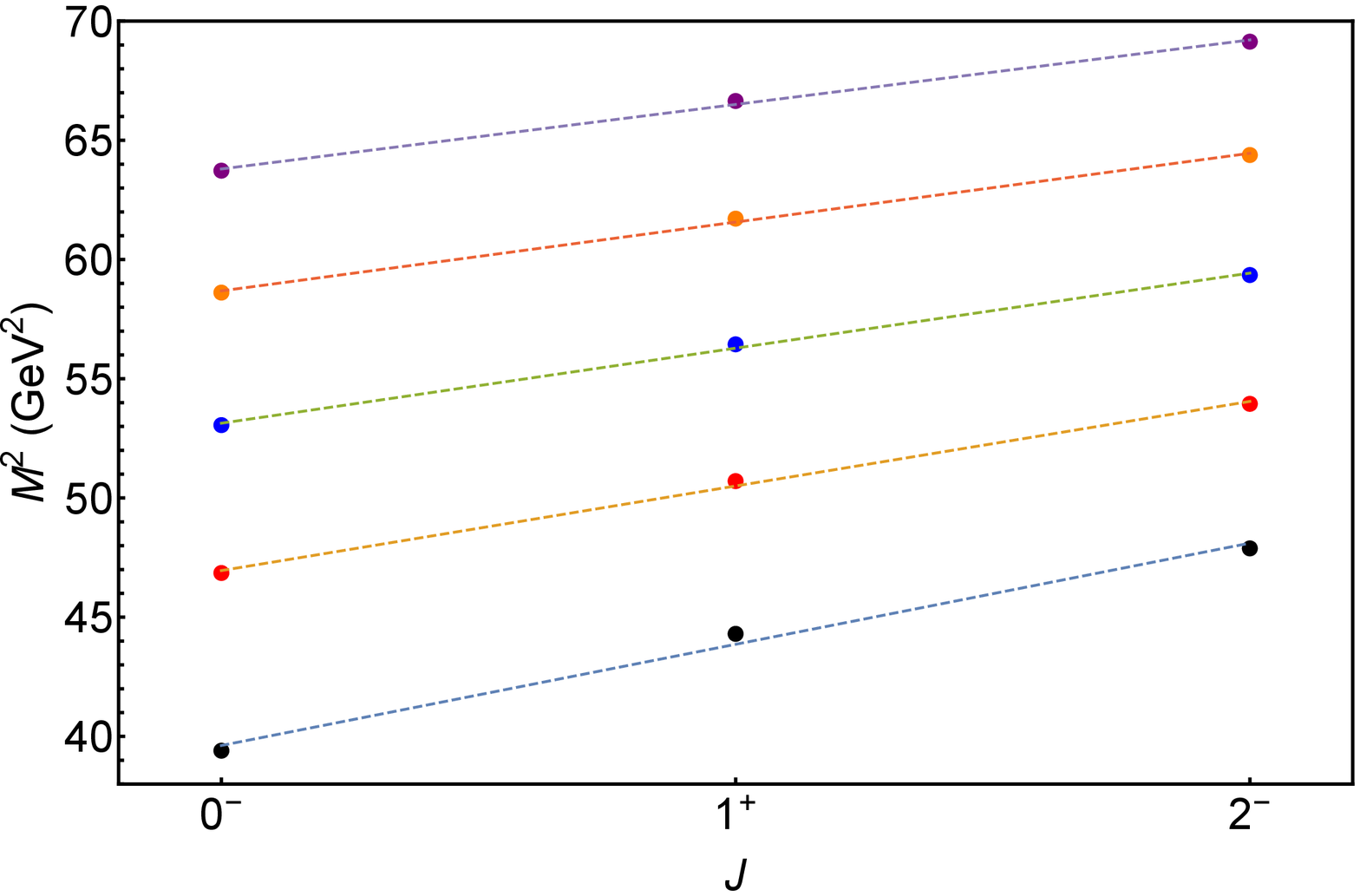} \\
\includegraphics[{width=0.6\textwidth}]{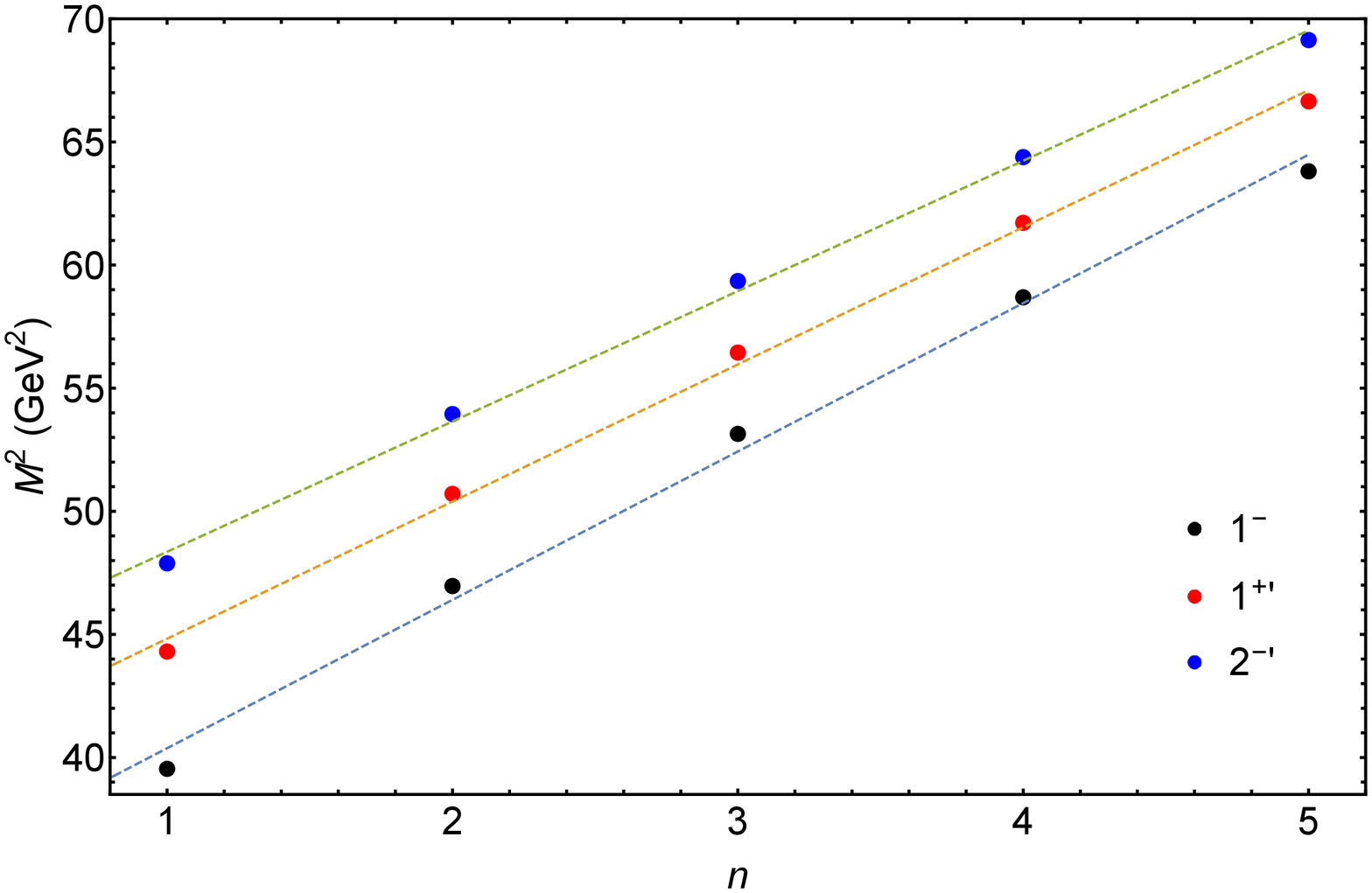} \\
\includegraphics[{width=0.6\textwidth}]{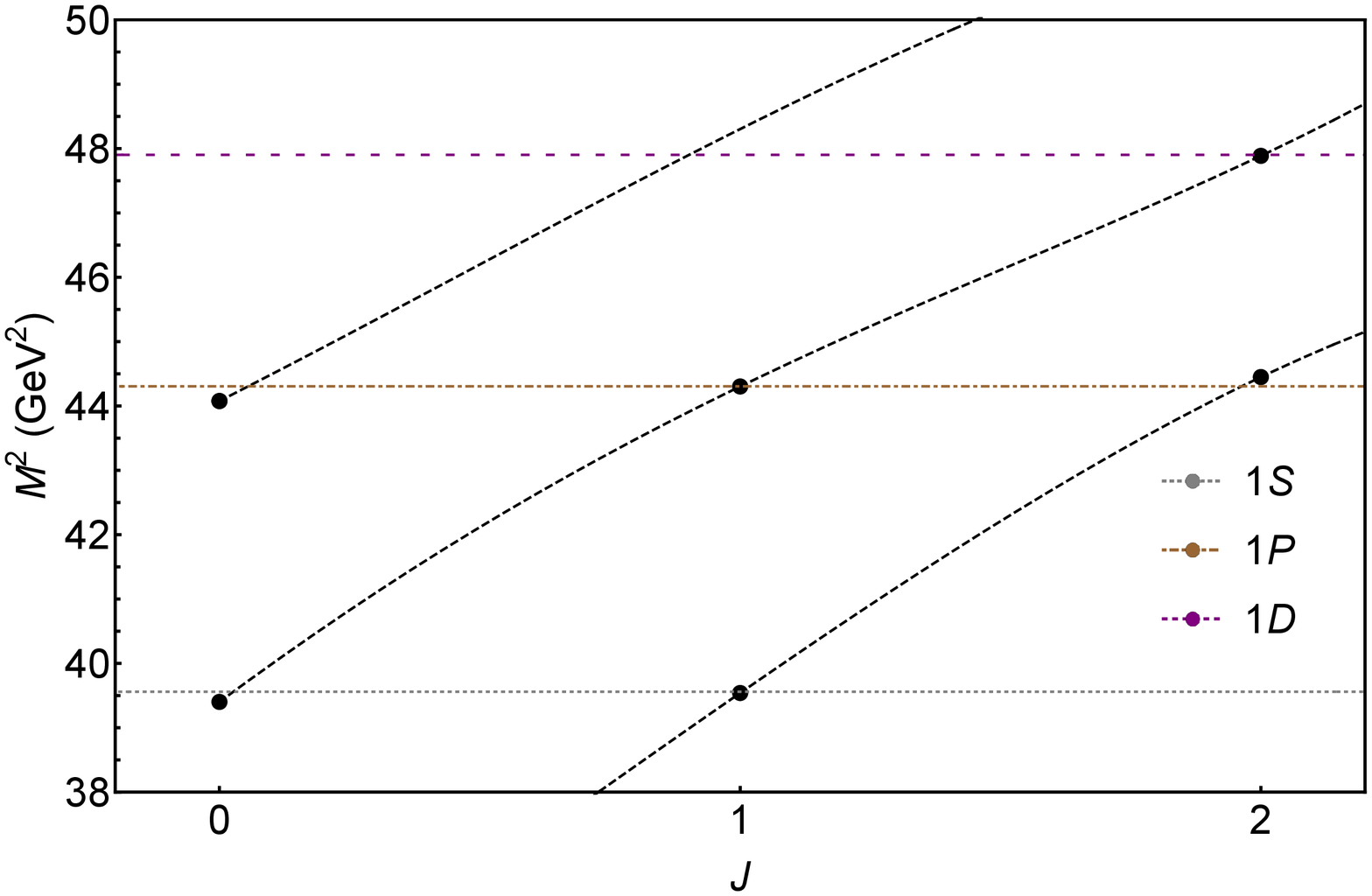} \\
\caption{Top Panel: Parent and Regge trajectories in $(J,M^2)$ plane for $B_c$ states with unnatural parity $P= (-1)^{J}$, for $n=1, ..., 5$ (from bottom to top).  Middle Panel: Regge trajectories in $(n,M^2)$ plane for $S$-wave vector, $P$-wave vector and $D$-wave tensor  $B_c$ states. Bottom Panel: Nonlinear  trajectories in $(J,M^2)$ plane starting from vector, pseudoscalar and scalar $B_c$ states (from bottom to top), with lines indicating the regions of $1S, 1P, 1D$ states. Circles represent the predicted masses shown in Table~\ref{TABLE-MESONS-BC1}, taking the values of the set of parameters III.}
\label{fig:Regge-Tr}
\end{figure}

In addition, the energy levels listed in Tables~\ref{TABLE-MESONS-BC1} and~\ref{TABLE-MESONS-BC2} allow us to obtain the mass relation between the ground states and their radial and angular excited states, and therefore construct the Regge trajectories in the $(n,M^2)$ and $(J,M^2)$ planes. They are then plotted in Fig.~\ref{fig:Regge-Tr}. 
In these plots, we assume that the Regge slopes are independent of charge conjugation, in
accordance with the $C$-invariance of QCD~\cite{Wei:2010zza}, and also that the slopes of the parity partner trajectories coincide. 

It can be remarked that the behavior of squared masses with radial quantum number and $J^P$ (top and middle panels) is not exactly linear. This fact is clearly pronounced in  $(n,M^2)$ plane, due to the high excitation number. More precisely, the daughter trajectories (incorporating both radially and orbitally excited states) manifest extrapolations closer to a linear fit. However, the parent trajectories (beginning from the ground states) reveal a nonlinear nature, mostly in the region of smaller mass. This is in qualitative accordance with other works that investigated heavy quarkonia states; see for instance Refs.~\cite{Wei:2010zza,Ebert:2011jc,Chen:2018hnx,Chen:2018bbr,Jia:2018vwl}.  Notwithstanding, using the  linear approximation for the Regge trajectories through the laws~\cite{Ebert:2011jc,Anisovich:2000kxa,Anisovich:2000ut,Afonin:2007aa},
\ben
M ^2 (J) & = &  \alpha_0  \beta +  J \,  \alpha, \nonumber \\
M ^2 (n) & = &  \beta_0 +  n \,  \beta, 
\label{LinearFit}
\een
where $\alpha_0, \beta_0$ are the intercepts and $\alpha, \beta$ the slopes of each corresponding trajectory on which the meson lies. Now applying this hypothesis in our scenario we can extract the parameters $\alpha_0, \beta_0, \alpha, \beta$ from the linear fits displayed in the top and middle panels of Fig.~\ref{fig:Regge-Tr}. The estimated values are listed in Table~\ref{tab:regge}. These results are in reasonable accordance with the existing literature, when compared for example with Ref.~\cite{Ebert:2011jc}.

\begin{longtable}{c|c|c}
  \caption{Fitted parameters for the linear fit in Eq.~(\ref{LinearFit}) of parent and daughter Regge trajectories in the top and middle panels of Fig.~\ref{fig:Regge-Tr}. The quantities are given in GeV${}^2$.}
%\vskip1.5mm
\label{tab:regge}
\\
\hline
\hline
\multicolumn{3}{c}{$(J,M^2)$ plane }\\
\hline
  Trajectory &  $\alpha_0$ &  $\alpha$ \\
\hline
 Parent & 39.620 & 4.243 \\
  1st Daughter & 46.956 & 3.548 \\
  2nd Daughter & 53.137 & 3.147 \\
  3rd Daughter & 58.689 & 2.885 \\ 
  4th Daughter & 63.801 & 2.705 \\
\hline
\hline
\multicolumn{3}{c}{$(n,M^2)$ plane} \\
\hline
  State &  $\beta_0$ &  $\beta$ \\
\hline
  $1^-$     & 34.349 & 6.027 \\
  $1^{+ \prime}$  & 39.253 & 5.570 \\
  $2^-$     & 43.060 & 5.294 \\
\hline
\hline
	     \end{longtable}
%\end{table}
%\end{widetext}

%%%%%%%%%%%%%%%%%%%%%%%%%%%%%%%%%%%%%%%%%%%%%%%%%%%%%%%%%%%%%%%%%%%%%%%%%%%%%%%%%%%%%%
%%%%%%%%%%%%%%%%%%%%%%%%%%%%%%%%%%%%%%%%%%%%%%%%%%%%%%%%%%%%%%%%%%%%%%%%%%%%%%%%%%%%%%
\subsection{Hyperfine splittings}
%%%%%%%%%%%%%%%%%%%%%%%%%%%%%%%%%%%%%%%%%%%%%%%%%%%%%%%%%%%%%%%%%%%%%%%%%%%%%%%%%%%%%%
%%%%%%%%%%%%%%%%%%%%%%%%%%%%%%%%%%%%%%%%%%%%%%%%%%%%%%%%%%%%%%%%%%%%%%%%%%%%%%%%%%%%%%

Additionally, another relevant feature to be noticed is the hyperfine splitting of $B_c$ states. We start with $S$-wave states. The hyperfine splittings $\Delta _{nS } ^{HFS}$ are listed in Table~\ref{split} for the set of parameters III chosen as well as other sets in order to see the influence of their change. Despite the small values obtained for these sets, as expected these splittings decrease for higher excited states, and get larger as the parameter $C_h$ grows. 
Our calculations yield the mass for $ B_c (2S)$ heavier than for $ B_c ^{\ast}(2S)$, which coincides with the other theoretical expectations. Nonetheless, our estimations engenders smaller $\Delta _{2S } ^{HFS} = M(2^3 S_1)-M(2^1 S_0)$ hyperfine splitting, as already remarked in Table~\ref{split}, but not too different from  the finding in Ref.~\cite{Li:2019tbn}.

\begin{longtable}{c|c|c|c|c|c}
  \caption{The hyperfine splittings for the $B_c$ system states, obtained for the set of parameters in Table~\ref{TABLE-QUARKS}. We use the respective definition for the hyperfine splitting: $\Delta _{nS } ^{HFS} =  M(n {}^3 S_1) - M(n {}^1 S_0)  $. The set used in Table~\ref{TABLE-MESONS-BC2} is in boldface. The quantities are given in MeV.}
%\vskip1.5mm
\label{split}
\\
\hline
\hline
 Set & $\Delta _{1S} ^{HFS}$ &  $\Delta _{2S} ^{HFS}$ & $\Delta _{3S} ^{HFS}$ & $\Delta _{4S} ^{HFS}$  & $\Delta _{5S} ^{HFS}$  \\
\hline
 I & 8 & 6 & 4 & 4 &  3  \\
 II  & 10 & 6 & 5  & 5 & 3  \\
 {\bf III} &  {\bf 11} &   {\bf 8} &  {\bf 6 } &  {\bf 5} &   {\bf 5}  \\
 IV  & 14 & 9 & 7 & 6 & 5 \\
\hline
\hline
	     \end{longtable}
%\end{table}
%\end{widetext}

Now we devote our attention to the hyperfine splitting of the $P$-wave states. For a systematic discussion of the hyperfine splitting for $P$-wave states in the context of hidden-flavor quarkonia, see for example Ref.~\cite{Peset:2018jkf}. In the case of bottom-charmed quarkonia, let us follow  as motivation the discussion done in Ref.~\cite{Chang:2019wpt}. Experimentally, it can be remarked that the spin-singlet $P$-wave states almost coincide with the spin-averaged centroid of the triplet~\cite{Tanabashi:2018oca} for $c\overline{c}$ and $b\overline{b}$ systems, yielding ideally 
\ben
 E(n^1 P_1) = \frac{1}{9}\left[ 5 E(n^3 P_2) + 3 E(n^3 P_1) + E(n^3 P_0)\right].
\label{PwaveM}
\een
Since for the $B_c$ system the $C$-parity is no longer a good quantum number, the states $n^3 P_1 - n^1 P_1$ can mix according to Eq.~(\ref{mix1}), and therefore Eq.~(\ref{PwaveM}) cannot be directly used. Nevertheless, assuming that the relation $E(n^3 P_1) \approx E(n^1 P_1) $ holds for $B_c$ mesons, then  Eq.~(\ref{PwaveM}) gives~\footnote{In Ref.~\cite{Chang:2019wpt} is argued that there should be possible hidden symmetry or an underlying principle that makes this relation valid for the $P$-level of charmonium and bottomonium systems, but also for the case of unequal-flavor case, like the the $B_c$ family.}
\ben
 E(n^3 P_0) + 5 E(n^3 P_2) = 3 \left[ E(n P_1) + E(n P_1 ^{\prime} )\right].
\label{PwaveM2}
\een
In order to test the validity of the relation above, the ratio among the masses can be introduced~\cite{Chang:2019wpt}:
\ben
r =  \frac{E(n^3 P_0) + 5 E(n^3 P_2)}{ 3 \left[ E(n P_1) + E(n P_1 ^{\prime} )\right]}. 
\label{PwaveRatio}
\een
So, the deviation from $r = 1$ accounts for how much Eq.~(\ref{PwaveM2}) is being violated. In this way, the estimations of $r$ for our calculations reported above are listed in Table~\ref{table:ratioPwave}. We see that the relative theoretical errors for the lowest-lying and radially-excited states are below 0.5\%, being even smaller for the excited states. Thus, the relation in Eq.~(\ref{PwaveM2}) holds to a fair precision. Another test that can be done is that when we augment the strength of the transverse potential, which is associated to the hyperfine interaction within the formalism used, the ratio $r$ increases (see for example the calculated masses for the set IV in Table~\ref{TABLE-MESONS-BC1}). This suggests that any deviation from $r=1$ depends on the hyperfine interaction, in consonance with the conclusions of Ref.~\cite{Chang:2019wpt}.

\begin{longtable}{c|c|c|c|c|c}
  \caption{Ratio $r$ defined in Eq.~(\ref{PwaveM2}) for lowest-lying and radially-excited $P$-wave $B_c$ states reported in Fig.~\ref{spectrum}.}
%\vskip1.5mm
\label{table:ratioPwave}
\\
\hline
\hline
 $n$ & 1 & 2 & 3 & 4 & 5 \\
\hline
 $ r $ & 1.0047 & 1.0031 & 1.0021 & 1.0014 & 1.0008 \\
\hline
\hline
	     \end{longtable}
%\end{table}
%\end{widetext}

As a final remark, this effective approach with a small number of parameters allows us construct the general aspects of   the bottom-charmed spectrum. Our predicted energy levels are in fair agreement with other predictions, providing a guide to the experimental search for the unobserved $B_c$ mesons.

%%%%%%%%%%%%%%%%%%%%%%%%%%%%%%%%%%%%%%%%%%%%%%%%%%%%%%%%%%%%%%%%%%%%%%%%%%%%%%%%%%%%%%
\section{Concluding Remarks}
\label{Conclusions}
%%%%%%%%%%%%%%%%%%%%%%%%%%%%%%%%%%%%%%%%%%%%%%%%%%%%%%%%%%%%%%%%%%%%%%%%%%%%%%%%%%%%%%

The purpose of this work has been the investigation of the $B_c$ meson spectrum by employing a different formalism with respect to the preceding analyses. The framework employed has been an effective version of the Coulomb gauge QCD and many-body techniques associated to the Tamm-Dancoff approximation (TDA). The interactions between quarks (quasiparticles) and antiquarks  (antiquasiparticles) have been given by the sum of an improved confining potential and a transverse hyperfine interaction with an Yukawa-type kernel, being interpreted as the exchange of a constituent gluon. 

Making use of a small number of parameters (dynamical mass of a constituent gluon $m_g $, current quark masses $m_{b}, m_c$ and the magnitude of the transverse potential $C_h$), this approach has allowed us to analyze the basic features of $B_c$ mesons. The calculated masses have been optimized in order to fit them to the  observed states by means of tuning of the parameters. Besides, the estimations of expected but yet-unobserved states are approximately in accordance with other findings in the literature using distinct formalisms. In particular, our calculations yield the mass for $ B_c (2S)$ lighter than for $ B_c ^{\ast}(2S)$, which coincides with the other theoretical expectations, but not with the CMS and LHCb results at the present moment. One expects that future findings will show that the true $ B_c ^{\ast}(2S)$ peak must be at a higher mass as the photon emitted in the $ B_c ^{\ast +}\rightarrow B_c ^{+} \gamma $ radiative transition can be reconstructed.

Another aspect regarded has been the mass relation between the ground states and their radial excited states in the $(n,M^2)$ and $(J,M^2)$ planes. The nonlinearity is more pronounced in the $(n,M^2)$ plane, due to the high excitation number used.  The parent trajectories (beginning from the ground states) reveal a nonlinear nature more evident than the daughter trajectories (incorporating both radially and orbitally excited states), which is in qualitative accordance with other works exploring quarkonia states and mesons.

Further, the hyperfine splitting of both $S$ and $P$-wave states has been studied. In both cases, we found that  these splittings decrease for higher excited states but become larger as the parameter $C_h$ grows, as expected, since it  drives the strength of the term associated to the hyperfine interaction.  In the case of $P$-wave states, the mass relation  in Eq.~(\ref{PwaveM2}) has relative theoretical errors for the lowest-lying and radially-excited states lower than 0.5\%. So, this framework engenders a reasonable precision for this mass relation involving the $P$-wave states.

Hence, we believe that this effective approach with a minimal number of parameters is capable of offering the general aspects of the bottom-charmed spectrum, in fair agreement with other predictions. At last,  all these works  provide a guide to the experimental search for the unobserved $B_c$ mesons. Some obvious extensions that deserve future studies are calculations and predictions on radiative and strong transitions and also on hybrid meson spectrum.

%%%%%%%%%%%%%%%%%%%%%%%%%%%%%%%%%%%%%%%%%%%%%%%%%%%%%%%%%%%%%%%%%%%%%%%%%%%%%%%%%%%%
%%%%%%%%%%%%%%%%%%%%%%%%%%%%%%%%%%%%%%%%%%%%%%%%%%%%%%%%%%%%%%%%%%%%%%%%%%%%%%%%%%%%
\begin{acknowledgements}
%%%%%%%%%%%%%%%%%%%%%%%%%%%%%%%%%%%%%%%%%%%%%%%%%%%%%%%%%%%%%%%%%%%%%%%%%%%%%%%%%%%%

We are grateful to Felipe J. Llanes-Estrada for support and discussions. We also would like to thank the Brazilian funding agencies for their financial support: CNPq (L.M.A.: contracts 308088/2017-4 and 400546/2016-7) and FAPESB (L.M.A.: contract INT0007/2016; F.M.C.J.: contract BOL2388/2017).

\end{acknowledgements} 
%%%%%%%%%%%%%%%%%%%%%%%%%%%%%%%%%%%%%%%%%%%%%%%%%%%%%%%%%%%%%%%%%%%%%%%%%%%%%%%%&&&&
%%%%%%%%%%%%%%%%%%%%%%%%%%%%%%%%%%%%%%%%%%%%%%%%%%%%%%%%%%%%%%%%%%%%%%%%%%%%%%%%%%%%

\appendix
%%%%%%%%%%%%%%%%%%%%%%%%%%%%%%%%%%%%%%%%%%%%%%%%%%%%%%%%%%%%%%%%%%%%%%%%%%%%%%%%%%%%%
%%%%%%%%%%%%%%%%%%%%%%%%%%%%%%%%%%%%%%%%%%%%%%%%%%%%%%%%%%%%%%%%%%%%%%%%%%%%%%%%%%%%%
\section{ Wave functions of TDA equation of motion} \label{sec:wave}
%%%%%%%%%%%%%%%%%%%%%%%%%%%%%%%%%%%%%%%%%%%%%%%%%%%%%%%%%%%%%%%%%%%%%%%%%%%%%%%%%%%%%
%%%%%%%%%%%%%%%%%%%%%%%%%%%%%%%%%%%%%%%%%%%%%%%%%%%%%%%%%%%%%%%%%%%%%%%%%%%%%%%%%%%%%

In this Appendix we present explicitly the $B_c$ meson spin-orbital wave functions, in terms of $L, S, J$  (helicity indices are not displayed): 
  \begin{itemize}
	\item pseudoscalar ($L=0; S= 0; J=0$),
\begin{eqnarray}
        \psi \left( 0^{-} \right)  & = &  \sqrt{\frac{1}{4 \pi}} \,\frac{i\sigma^2}{\sqrt{2}}, 
	\label{WF_PS}
\end{eqnarray}
	
	\item scalar ($L=1; S= 1; J=0$),
\begin{eqnarray}
         \psi  \left( 0^{+} \right)  & = &  - \sqrt{\frac{1}{4 \pi}}   \, \left(\boldsymbol{\sigma}\cdot\boldsymbol{\hat{k}}\right)\,\frac{i\sigma^2}{\sqrt{2}}, 
	\label{WF_S}
\end{eqnarray}
	
	\item vector ($L=0; S= 1; J=1$),
\begin{eqnarray}
        \psi \left( 1^{-} \right)  & = &   \sqrt{\frac{3}{4 \pi}}  \, \boldsymbol{\sigma} \,\frac{i\sigma^2}{\sqrt{2}}, 
	\label{WF_V}
\end{eqnarray}

	\item axial ($L=1; S= 0; J=1$),
\begin{eqnarray}
        \psi  \left( 1^{+} \right)  & = &   i \sqrt{\frac{3}{4 \pi}}  \, \boldsymbol{\hat{k}} \,\frac{i\sigma^2}{\sqrt{2}}, 
	\label{WF_PVM}
\end{eqnarray}
		
	\item axial ($L=1; S= 1; J=1$),
\begin{eqnarray}
        \psi \left( 1^{+ \, \prime} \right)  & = &  - i \sqrt{\frac{3}{8 \pi}}  \, \left(\boldsymbol{\sigma}\times\boldsymbol{\hat{k}}\right) \,\frac{i\sigma^2}{\sqrt{2}}, 
	\label{WF_PVP}
\end{eqnarray}

	\item Tensor ($L=1; S= 1; J=2$),
\begin{eqnarray}
        \psi \left( 2^{+} \right)  & = &    \sqrt{\frac{3}{4 \pi}}  \, \boldsymbol{\sigma} \, \boldsymbol{\hat{k}} \,\frac{i\sigma^2}{\sqrt{2}}, 
	\label{WF_T}
\end{eqnarray}

	\item Pseudotensor ($L=2; S= 0; J=2$),
\begin{eqnarray}
        \psi \left( 2^{-} \right)  & = &    \sqrt{\frac{5}{4 \pi}}  \, \boldsymbol{\hat{k}}   \, \boldsymbol{\hat{k}} \,\frac{i\sigma^2}{\sqrt{2}}, 
	\label{WF_PTP}
\end{eqnarray}

	\item Pseudotensor ($L=2; S= 1; J=2$),
\begin{eqnarray}
        \psi \left( 2^{- \, \prime} \right)  & = & -   \sqrt{\frac{5}{4 \pi}}  \, \left(\boldsymbol{\sigma}\times\boldsymbol{\hat{k}}\right)  \, \boldsymbol{\hat{k}} \,\frac{i\sigma^2}{\sqrt{2}}, 
	\label{WF_PTM}
\end{eqnarray}

\end{itemize}
The factor $(i\sigma^2)$ is introduced in order to use the same convention of Ref.~\cite{Abreu:2019adi}.

%%%%%%%%%%%%%%%%%%%%%%%%%%%%%%%%%%%%%%%%%%%%%%%%%%%%%%%%%%%%%%%%%%%%%%%%%%%%%%%%%%%%%
%%%%%%%%%%%%%%%%%%%%%%%%%%%%%%%%%%%%%%%%%%%%%%%%%%%%%%%%%%%%%%%%%%%%%%%%%%%%%%%%%%%%%
\section{ Kernels of TDA equation of motion} \label{sec:kernels}
%%%%%%%%%%%%%%%%%%%%%%%%%%%%%%%%%%%%%%%%%%%%%%%%%%%%%%%%%%%%%%%%%%%%%%%%%%%%%%%%%%%%%
%%%%%%%%%%%%%%%%%%%%%%%%%%%%%%%%%%%%%%%%%%%%%%%%%%%%%%%%%%%%%%%%%%%%%%%%%%%%%%%%%%%%%

Here we set out the relevant kernels obtained for the mesons described by the wavefunctions given by Eqs.~(\ref{WF_PS})-(\ref{WF_PTM}): 

  \begin{itemize}
	\item pseudoscalar ($L=0; S= 0; J=0$),
\begin{eqnarray}
        K^{\left( 0^{-} \right)}\left(k, q\right)  & = &   V_{1} \left(a_{5} + a_{6}\right) + V_{0} \left(a_{7} + a_{8}\right) \nonumber \\			
        & & + 2 U_{0} \left(a_{1} + a_{2}\right)  
		- 2 W_{0} \left(a_{3} + a_{4}\right);
	\label{K_PS}
\end{eqnarray}
	
	\item scalar ($L=1; S= 1; J=0$),
\begin{eqnarray}
        K^{\left( 0^{+} \right)}\left(k, q\right)  & = &   V_{0} \left(a_{5} + a_{6}\right) + V_{1} \left(a_{7} + a_{8}\right) - 2 U_{0} \left(a_{3} + a_{4}\right)  			\nonumber \\			& & 			
        + \left(  U_1 + W_{0} - k q Z_{0} \right) \left(a_{1} + a_{2}\right);
	\label{K_PS}
\end{eqnarray}
	
	\item vector ($L=0; S= 1; J=1$),
\begin{eqnarray}
			K^{\left( 1^{-} \right)}\left(k, q\right)  & = &   \frac{1}{3} \left[3 V_{1} \left(a_{5} + a_{6}\right) + a_{8} \left(4 V_{2} - V_{0}\right) + 3 a_{7} V_{0} - 2 \left(a_{1} + a_{2}\right) U_{0} +\right. \nonumber \\
			& & \left.+ 2 \left(a_{3} + a_{4}\right) U_{1} + 2 q k \left(a_{3} + a_{4}\right) Z_{0} + 4 \left(a_{1} k^{2} + a_{2} q^{2}\right) Z_{0} \right];
	\label{K_V}
\end{eqnarray}
		
	\item axial ($L=1; S= 0; J=1$),
\begin{eqnarray}
			K^{\left( 1^{+} \right)} \left(k, q\right)  & = &   \left(a_{5} + a_{6}\right) V_{2} + \left(a_{7} + a_{8}\right) V_{1} + 2 \left(a_{1} + a_{2}\right) U_{1}
		%	 +\right.\\
		%	 & & \left.
			 - 2 \left(a_{3} + a_{4}\right) W_{1};
	\label{K_PV1}
\end{eqnarray}
		
	\item axial ($L=1; S= 1; J=1$),
\begin{eqnarray}
			K^{\left( 1^{+ \, \prime} \right)}\left(k, q\right)  & = &   \frac{1}{2} \left(V_{0} + V_{2}\right) \left(a_{5} + a_{6}\right) + \frac{1}{2} \left(U_{0} + U_{2} - 2 W_{1}\right) \left(a_{3} + a_{4}\right) \nonumber  \\
			& & + V_{1} \left(a_{7} + a_{8}\right) + Z_{1} \left(a_{1} k^{2} + a_{2} q^{2}\right) + Z_{0} \frac{1}{2} \left(k^{2} - q^{2}\right) \left(a_{4} - a_{3}\right).
	\label{K_PV2}
\end{eqnarray}

	\item Tensor ($L=1; S= 1; J=2$),
\begin{eqnarray}
			K^{\left( 2^{+} \right)}\left(k, q\right)  & = &   \frac{1}{2} \left(3 V_{2} - V_{0}\right) \left(a_{5} + a_{6}\right) + V_{1} a_{7} + \frac{1}{5}(12 V_3 - 7 V_1 ) a_8
\nonumber  \\
			& & + \frac{1}{5}\left(  U_{1} - 5 W_0 - 10 k q Z_0 \right) \left(a_{1} + a_{2}\right) 
			+ \frac{12}{5} Z_{1} \left(a_{1} k^{2} + a_{2} q^{2}\right) 
			\nonumber  \\
			& & + \frac{1}{10} \left(27 
			U_2 - 15 W_1 - 9 k q Z_1 - 4 U_0 \right) \left(a_{3} + a_{4}\right); 
	\label{K_T1}
\end{eqnarray}

	\item Pseudotensor ($L=2; S= 0; J=2$),
\begin{eqnarray}
			K^{\left( 2^{- } \right)}\left(k, q\right)  & = &   \frac{1}{2} \left(3 V_{3} - V_{1}\right) \left(a_{5} + a_{6}\right) +\frac{1}{2} \left(3 V_{2} - V_{0}\right) \left(a_{7} + a_{8}\right) 
\nonumber  \\
			& & + \left( 3 U_{2} - U_0 \right) \left(a_{1} + a_{2}\right) 
			+ \left( 3 W_{2} - W_0 \right) \left(a_{3} + a_{4}\right) ; 
	\label{K_T2}
\end{eqnarray}

	\item Pseudotensor ($L=2; S= 1; J=2$),
\begin{eqnarray}
			K^{\left( 2^{- \, \prime} \right)}\left(k, q\right)  & = &   V_{3} \left(a_{5} + a_{6}\right) +\frac{1}{2} \left( 3 V_{2} - V_{0}\right) \left(a_{7} + a_{8}\right) 
\nonumber  \\
			& & + k q Z_1 \left(a_{1} + a_{2}\right) 
			+ \frac{1}{2} \left( 2 Z_{2} - Z_0 \right) \left(a_{3} + a_{4}\right) ; 
	\label{K_T3}
\end{eqnarray}

\end{itemize}
where the coefficients $a_i$ are given by
\begin{eqnarray}
	a_{1} &= & \sqrt{1 + s_{k}^b } \sqrt{1 + s_{k}^{c} } \sqrt{1 - s_{q}^b } \sqrt{1 - s_{q}^{c} },\label{eq:d17} \nonumber \\
	a_{2} &= & \sqrt{1 - s_{k}^b } \sqrt{1 - s_{k}^{c} } \sqrt{1 + s_{q}^b } \sqrt{1 + s_{q}^{c} }, \nonumber \\
	a_{3} &= &\sqrt{1 + s_{k}^b } \sqrt{1 - s_{k}^{c} } \sqrt{1 - s_{q}^b } \sqrt{1 + s_{q}^{c} }, \nonumber \\
	a_{4} &= & \sqrt{1 - s_{k}^b} \sqrt{1 + s_{k}^{c} } \sqrt{1 + s_{q}^b} \sqrt{1 - s_{q}^{c} }, \nonumber \\
	a_{5} &= & \sqrt{1 + s_{k}^b } \sqrt{1 - s_{k}^{c} } \sqrt{1 + s_{q}^b } \sqrt{1 - s_{q}^{c} }, \nonumber \\
	a_{6} &= & \sqrt{1 - s_{k}^b } \sqrt{1 + s_{k}^{c} } \sqrt{1 - s_{q}^b } \sqrt{1 + s_{q}^{c} }, \nonumber \\
	a_{7} &= & \sqrt{1 + s_{k}^b } \sqrt{1 +  s_{k}^{c} } \sqrt{1 + s_{q}^b } \sqrt{1 + s_{q}^{c} }, \nonumber \\
	a_{8} &= & \sqrt{1 - s_{k}^b } \sqrt{1 - s_{k}^{c} } \sqrt{1 - s_{q}^b } \sqrt{1 - s_{q}^{c} }.
	\label{a_coeff}
\end{eqnarray}
The functions $s_{k(q)}^{b} $ and $s_{k(q)}^{c} $ are dependent of the respective gap angle obtained by solving the gap equation for the $b$ and $c$ quarks, respectively. Besides the functions $V_n$, $U_n$ and $W_n$, defined in Eqs.~(\ref{ang_int}) and ~(\ref{W_func}), the auxiliary $Z$-function has been also introduced: 
\be 
Z(|\mathbf{k} - \mathbf{q}|) \equiv U(|\mathbf{k} - \mathbf{q}|) \frac{ 1 - x^2 }{|\mathbf{k} - \mathbf{q}|^2}. 
\label{Z_func}
\ee

%%%%%%%%%%%%%%%%%%%%%%%%%%%%%%%%%%%%%%%%%%%%%%%%%%%%%%%%%%%%%%%%%%%%%%%%%%%%%%%%%%%%%
%%%%%%%%%%%%%%%%%%%%%%%%%%%%%%%%%%%%%%%%%%%%%%%%%%%%%%%%%%%%%%%%%%%%%%%%%%%%%%%%%%%%%


\begin{thebibliography}{99}
%%%%%%%%%%%%%%%%%%%%%%%%%%%%%%%%%%%%%%%%%%%%%%%%%%%%%%%%%%%%%%%%%%%%%%%%%%%%%%%%%%%%%
%%%%%%%%%%%%%%%%%%%%%%%%%%%%%%%%%%%%%%%%%%%%%%%%%%%%%%%%%%%%%%%%%%%%%%%%%%%%%%%%%%%%%


%\cite{Abe:1998wi}
\bibitem{Abe:1998wi}
F.~Abe \textit{et al.} [CDF],
%``Observation of the $B_c$ meson in $p\bar{p}$ collisions at $\sqrt{s} = 1.8$ TeV,''
Phys. Rev. Lett. \textbf{81}, 2432-2437 (1998)
doi:10.1103/PhysRevLett.81.2432
[arXiv:hep-ex/9805034 [hep-ex]].
%388 citations counted in INSPIRE as of 14 May 2020

%\cite{Abazov:2008kv}
\bibitem{Abazov:2008kv}
V.~Abazov \textit{et al.} [D0],
%``Observation of the $B_c$ Meson in the Exclusive Decay $B_c \to J/\psi \pi$,''
Phys. Rev. Lett. \textbf{101}, 012001 (2008)
doi:10.1103/PhysRevLett.101.012001
[arXiv:0802.4258 [hep-ex]].
%120 citations counted in INSPIRE as of 14 May 2020

%\cite{Aaij:2012dd}
\bibitem{Aaij:2012dd}
R.~Aaij \textit{et al.} [LHCb],
%``Measurements of $B_c^+$ production and mass with the $B_c^+ \to J/\psi \pi^+$ decay,''
Phys. Rev. Lett. \textbf{109}, 232001 (2012)
doi:10.1103/PhysRevLett.109.232001
[arXiv:1209.5634 [hep-ex]].
%94 citations counted in INSPIRE as of 14 May 2020

%\cite{Tanabashi:2018oca}
\bibitem{Tanabashi:2018oca} 
  M.~Tanabashi {\it et al.} [Particle Data Group],
  %``Review of Particle Physics,''
  Phys.\ Rev.\ D {\bf 98}, no. 3, 030001 (2018).
  doi:10.1103/PhysRevD.98.030001
  %%CITATION = doi:10.1103/PhysRevD.98.030001;%%
  %1003 citations counted in INSPIRE as of 14 Feb 2019


%\cite{Aad:2014laa}
\bibitem{Aad:2014laa}
G.~Aad \textit{et al.} [ATLAS],
%``Observation of an Excited $B_c^\pm$ Meson State with the ATLAS Detector,''
Phys. Rev. Lett. \textbf{113}, no.21, 212004 (2014)
doi:10.1103/PhysRevLett.113.212004
[arXiv:1407.1032 [hep-ex]].
%87 citations counted in INSPIRE as of 06 May 2020

%\cite{Sirunyan:2019osb}
\bibitem{Sirunyan:2019osb}
A.~M.~Sirunyan \textit{et al.} [CMS],
%``Observation of Two Excited B$^+_\mathrm{c}$ States and Measurement of the B$^+_\mathrm{c}$(2S) Mass in pp Collisions at $\sqrt{s} =$ 13 TeV,''
Phys. Rev. Lett. \textbf{122}, no.13, 132001 (2019)
doi:10.1103/PhysRevLett.122.132001
[arXiv:1902.00571 [hep-ex]].
%34 citations counted in INSPIRE as of 29 Apr 2020

%\cite{Aaij:2019ldo}
\bibitem{Aaij:2019ldo}
R.~Aaij \textit{et al.} [LHCb],
%``Observation of an excited $B_c^+$ state,''
Phys. Rev. Lett. \textbf{122}, no.23, 232001 (2019)
doi:10.1103/PhysRevLett.122.232001
[arXiv:1904.00081 [hep-ex]].
%21 citations counted in INSPIRE as of 07 May 2020


%\cite{Eichten:1994gt}
\bibitem{Eichten:1994gt}
E.~J.~Eichten and C.~Quigg,
%``Mesons with beauty and charm: Spectroscopy,''
Phys. Rev. D \textbf{49}, 5845-5856 (1994)
doi:10.1103/PhysRevD.49.5845
[arXiv:hep-ph/9402210 [hep-ph]].
%484 citations counted in INSPIRE as of 22 Apr 2020


%\cite{Monteiro:2016ijw}
\bibitem{Monteiro:2016ijw}
A.~P.~Monteiro, M.~Bhat and K.~B.~Vijaya~Kumar,
%``Mass spectra and decays of ground and orbitally excited $c\bar{b}$ states in nonrelativistic quark model,''
Int. J. Mod. Phys. A \textbf{32}, no.04, 1750021 (2017)
doi:10.1142/S0217751X1750021X
[arXiv:1607.07594 [hep-ph]].
%11 citations counted in INSPIRE as of 05 May 2020


%\cite{Monteiro:2016rzi}
\bibitem{Monteiro:2016rzi}
A.~P.~Monteiro, M.~Bhat and K.~B.~Vijaya~Kumar,
%``$c\bar{b}$ spectrum and decay properties with coupled channel effects,''
Phys. Rev. D \textbf{95}, no.5, 054016 (2017)
doi:10.1103/PhysRevD.95.054016
[arXiv:1608.05782 [hep-ph]].
%13 citations counted in INSPIRE as of 05 May 2020

\bibitem{Soni:2017wvy}
N.~R.~Soni, B.~R.~Joshi, R.~P.~Shah, H.~R.~Chauhan and J.~N.~Pandya,
%``$Q\bar{Q}$ ( $Q\in \{b, c\}$ ) spectroscopy using the Cornell potential,''
Eur. Phys. J. C \textbf{78}, no.7, 592 (2018)
doi:10.1140/epjc/s10052-018-6068-6
[arXiv:1707.07144 [hep-ph]].
%28 citations counted in INSPIRE as of 20 Jul 2020


%\cite{Akbar:2018hiw}
\bibitem{Akbar:2018hiw}
N.~Akbar, M.~Atif Sultan, B.~Masud and F.~Akram,
%``Conventional and hybrid B$_{c}$ mesons in an extended potential model,''
Eur. Phys. J. A \textbf{55}, no.5, 82 (2019)
doi:10.1140/epja/i2019-12735-1
[arXiv:1811.07552 [hep-ph]].
%4 citations counted in INSPIRE as of 22 Apr 2020

%\cite{Eichten:2019gig}
\bibitem{Eichten:2019gig}
E.~J.~Eichten and C.~Quigg,
%``Mesons with Beauty and Charm: New Horizons in Spectroscopy,''
Phys. Rev. D \textbf{99}, no.5, 054025 (2019)
doi:10.1103/PhysRevD.99.054025
[arXiv:1902.09735 [hep-ph]].
%19 citations counted in INSPIRE as of 22 Apr 2020

%\cite{Li:2019tbn}
\bibitem{Li:2019tbn}
Q.~Li, M.~S.~Liu, L.~S.~Lu, Q.~F.~Lu, L.~C.~Gui and X.~H.~Zhong,
%``Excited bottom-charmed mesons in a nonrelativistic quark model,''
Phys. Rev. D \textbf{99}, no.9, 096020 (2019)
doi:10.1103/PhysRevD.99.096020
[arXiv:1903.11927 [hep-ph]].
%12 citations counted in INSPIRE as of 22 Apr 2020

%\cite{Chang:2019wpt}
\bibitem{Chang:2019wpt}
L.~Chang, M.~Chen, X.~q.~Li, Y.~x.~Liu and K.~Raya,
%``Can the Miraculous Mass Relation in the $P$-wave Spectroscopy of Charmonia and Bottomonia be extended to $B_{c}$ Mesons,''
[arXiv:1912.08339 [nucl-th]].
%1 citations counted in INSPIRE as of 15 May 2020

%\cite{Ortega:2020uvc}
\bibitem{Ortega:2020uvc}
P.~G.~Ortega, J.~Segovia, D.~R.~Entem and F.~Fernandez,
%``Spectroscopy of $\mathbf{B_c}$ mesons and the possibility of finding exotic $\mathbf{B_c}$-like structures,''
Eur. Phys. J. C \textbf{80}, no.3, 223 (2020)
doi:10.1140/epjc/s10052-020-7764-6
[arXiv:2001.08093 [hep-ph]].
%0 citations counted in INSPIRE as of 22 Apr 2020



%\cite{Godfrey:1985xj}
\bibitem{Godfrey:1985xj} 
  S.~Godfrey and N.~Isgur,
  %``Mesons in a Relativized Quark Model with Chromodynamics,''
  Phys.\ Rev.\ D {\bf 32}, 189 (1985).
  doi:10.1103/PhysRevD.32.189
  %%CITATION = doi:10.1103/PhysRevD.32.189;%%
  %2708 citations counted in INSPIRE as of 27 Mar 2020


%\cite{Zeng:1994vj}
\bibitem{Zeng:1994vj}
J.~Zeng, J.~W.~Van Orden and W.~Roberts,
%``Heavy mesons in a relativistic model,''
Phys. Rev. D \textbf{52}, 5229-5241 (1995)
doi:10.1103/PhysRevD.52.5229
[arXiv:hep-ph/9412269 [hep-ph]].
%156 citations counted in INSPIRE as of 30 Apr 2020

%\cite{Gupta:1995ps}
\bibitem{Gupta:1995ps}
S.~N.~Gupta and J.~M.~Johnson,
%``B(c) spectroscopy in a quantum chromodynamic potential model,''
Phys. Rev. D \textbf{53}, 312-314 (1996)
doi:10.1103/PhysRevD.53.312
[arXiv:hep-ph/9511267 [hep-ph]].
%65 citations counted in INSPIRE as of 22 Apr 2020


%\cite{Ebert:2002pp}
\bibitem{Ebert:2002pp}
D.~Ebert, R.~N.~Faustov and V.~O.~Galkin,
%``Properties of heavy quarkonia and $B_c$ mesons in the relativistic quark model,''
Phys. Rev. D \textbf{67}, 014027 (2003)
doi:10.1103/PhysRevD.67.014027
[arXiv:hep-ph/0210381 [hep-ph]].
%412 citations counted in INSPIRE as of 15 May 2020

%\cite{Ikhdair:2003ry}
\bibitem{Ikhdair:2003ry}
S.~M.~Ikhdair and R.~Sever,
%``Spectroscopy of $B_c$ meson in a semirelativistic quark model using the shifted large $N$ expansion method,''
Int. J. Mod. Phys. A \textbf{19}, 1771-1792 (2004)
doi:10.1142/S0217751X0401780X
[arXiv:hep-ph/0310295 [hep-ph]].
%41 citations counted in INSPIRE as of 22 Apr 2020

%\cite{Godfrey:2004ya}
\bibitem{Godfrey:2004ya}
S.~Godfrey,
%``Spectroscopy of $B_c$ mesons in the relativized quark model,''
Phys. Rev. D \textbf{70}, 054017 (2004)
doi:10.1103/PhysRevD.70.054017
[arXiv:hep-ph/0406228 [hep-ph]].
%155 citations counted in INSPIRE as of 22 Apr 2020

\bibitem{Kiselev:1994rc}
S.~S.~Gershtein, V.~V.~Kiselev, A.~K.~Likhoded and A.~V.~Tkabladze,
%``B(c) spectroscopy,''
Phys. Rev. D \textbf{51}, 3613-3627 (1995)
doi:10.1103/PhysRevD.51.3613
[arXiv:hep-ph/9406339 [hep-ph]].

%\cite{Aliev:2019wcm}
\bibitem{Aliev:2019wcm}
T.~Aliev, T.~Barakat and S.~Bilmis,
%``Properties of excited $B_c$ states in QCD,''
doi:10.1016/j.nuclphysb.2019.114726
[arXiv:1905.11750 [hep-ph]].
%0 citations counted in INSPIRE as of 22 Apr 2020



%\cite{Wang:2012kw}
\bibitem{Wang:2012kw}
Z.~G.~Wang,
%``Analysis of the vector and axialvector $B_c$ mesons with QCD sum rules,''
Eur. Phys. J. A \textbf{49}, 131 (2013)
doi:10.1140/epja/i2013-13131-7
[arXiv:1203.6252 [hep-ph]].
%23 citations counted in INSPIRE as of 14 May 2020

%\cite{Allison:2004be}
\bibitem{Allison:2004be}
I.~F.~Allison, C.~T.~H.~Davies, A.~Gray, A.~S.~Kronfeld, P.~B.~Mackenzie, J.~N.~Simone,
%``Mass of the $B_c$ meson in three-flavor lattice QCD,''
Phys. Rev. Lett. \textbf{94}, 172001 (2005)
doi:10.1103/PhysRevLett.94.172001
[arXiv:hep-lat/0411027 [hep-lat]].
%142 citations counted in INSPIRE as of 22 Apr 2020

%\cite{Dowdall:2012ab}
\bibitem{Dowdall:2012ab}
R.~J.~Dowdall, C.~T.~H.~Davies, T.~C.~Hammant and R.~R.~Horgan,
%``Precise heavy-light meson masses and hyperfine splittings from lattice QCD including charm quarks in the sea,''
Phys. Rev. D \textbf{86}, 094510 (2012)
doi:10.1103/PhysRevD.86.094510
[arXiv:1207.5149 [hep-lat]].
%113 citations counted in INSPIRE as of 14 May 2020

%\cite{Mathur:2018epb}
\bibitem{Mathur:2018epb}
N.~Mathur, M.~Padmanath and S.~Mondal,
%``Precise predictions of charmed-bottom hadrons from lattice QCD,''
Phys. Rev. Lett. \textbf{121}, no.20, 202002 (2018)
doi:10.1103/PhysRevLett.121.202002
[arXiv:1806.04151 [hep-lat]].
%37 citations counted in INSPIRE as of 22 Apr 2020

%\cite{Chen:2020ecu}
\bibitem{Chen:2020ecu}
M.~Chen, L.~Chang and Y.~X.~Liu,
%``$B_c$ meson spectrum via Dyson-Schwinger equation and Bethe-Salpeter equation approach,''
Phys. Rev. D \textbf{101}, no.5, 056002 (2020)
doi:10.1103/PhysRevD.101.056002
[arXiv:2001.00161 [hep-ph]].
%2 citations counted in INSPIRE as of 22 Apr 2020

%\cite{Szczepaniak:1995cw}
\bibitem{Szczepaniak:1995cw}
A.~Szczepaniak, E.~S.~Swanson, C.~R.~Ji and S.~R.~Cotanch,
%``Glueball spectroscopy in a relativistic many body approach to hadron structure,''
Phys. Rev. Lett. \textbf{76}, 2011-2014 (1996)
doi:10.1103/PhysRevLett.76.2011
[arXiv:hep-ph/9511422 [hep-ph]].
%147 citations counted in INSPIRE as of 15 May 2020

%\cite{Cotanch:1998ph}
\bibitem{Cotanch:1998ph}
S.~Cotanch, A.~Szczepaniak, E.~Swanson and C.~Ji,
%``QCD Hamiltonian approach for the glueball spectrum,''
Nucl. Phys. A \textbf{631}, 640C-643C (1998)
doi:10.1016/S0375-9474(98)00082-7
%7 citations counted in INSPIRE as of 15 May 2020


%\cite{LlanesEstrada:1999uh}
\bibitem{LlanesEstrada:1999uh} 
  F.~J.~Llanes-Estrada and S.~R.~Cotanch,
  %``Meson structure in a relativistic many body approach,''
  Phys.\ Rev.\ Lett.\  {\bf 84}, 1102 (2000)
  doi:10.1103/PhysRevLett.84.1102.
  %%CITATION = doi:10.1103/PhysRevLett.84.1102;%%

%\cite{Szczepaniak:2001rg}
\bibitem{Szczepaniak:2001rg} 
  A.~P.~Szczepaniak and E.~S.~Swanson,
  %``Coulomb gauge QCD, confinement, and the constituent representation,''
  Phys.\ Rev.\ D {\bf 65}, 025012 (2001)
  doi:10.1103/PhysRevD.65.025012.
  %%CITATION = doi:10.1103/PhysRevD.65.025012;%%


%\cite{LlanesEstrada:2001kr}
\bibitem{LlanesEstrada:2001kr} 
  F.~J.~Llanes-Estrada and S.~R.~Cotanch,
  %``Relativistic many body Hamiltonian approach to mesons,''
  Nucl.\ Phys.\ A {\bf 697}, 303 (2002)
  doi:10.1016/S0375-9474(01)01237-4
  [hep-ph/0101078].
  %%CITATION = doi:10.1016/S0375-9474(01)01237-4;%%

%\cite{Ligterink:2003hd}
\bibitem{Ligterink:2003hd} 
  N.~Ligterink and E.~S.~Swanson,
  %``A Coulomb gauge model of mesons,''
  Phys.\ Rev.\ C {\bf 69}, 025204 (2004)
  doi:10.1103/PhysRevC.69.025204
  [hep-ph/0310070].
  %%CITATION = doi:10.1103/PhysRevC.69.025204;%%

\bibitem{LlanesEstrada:2004wr} 
  F.~J.~Llanes-Estrada, S.~R.~Cotanch, A.~P.~Szczepaniak and E.~S.~Swanson,
  %``Hyperfine meson splittings: Chiral symmetry versus transverse gluon exchange,''
  Phys.\ Rev.\ C {\bf 70}, 035202 (2004)
  doi:10.1103/PhysRevC.70.035202
  [hep-ph/0402253].
  %%CITATION = doi:10.1103/PhysRevC.70.035202;%%



%\cite{LlanesEstrada:2005jf}
\bibitem{LlanesEstrada:2005jf} 
  F.~J.~Llanes-Estrada, P.~Bicudo and S.~R.~Cotanch,
  %``Oddballs and a low odderon intercept,''
  Phys.\ Rev.\ Lett.\  {\bf 96}, 081601 (2006)
  doi:10.1103/PhysRevLett.96.081601.
  %%CITATION = doi:10.1103/PhysRevLett.96.081601;%%

%\cite{Szczepaniak:2005xi}
\bibitem{Szczepaniak:2005xi} 
  A.~P.~Szczepaniak and P.~Krupinski,
  %``Coulomb energy and gluon distribution in the presence of static sources,''
  Phys.\ Rev.\ D {\bf 73}, 034022 (2006)
  doi:10.1103/PhysRevD.73.034022 .
  %%CITATION = doi:10.1103/PhysRevD.73.034022;%%

%\cite{General:2007bk}
\bibitem{General:2007bk}
I.~J.~General, P.~Wang, S.~R.~Cotanch and F.~J.~Llanes-Estrada,
%``Light 1-+ exotics: Molecular resonances,''
Phys. Lett. B \textbf{653}, 216-223 (2007)
doi:10.1016/j.physletb.2007.08.015
[arXiv:0707.1286 [hep-ph]].
%16 citations counted in INSPIRE as of 15 May 2020

%\cite{Guo:2007sm}
\bibitem{Guo:2007sm}
P.~Guo, A.~P.~Szczepaniak, G.~Galata, A.~Vassallo and E.~Santopinto,
%``Gluelump spectrum from Coulomb gauge QCD,''
Phys. Rev. D \textbf{77}, 056005 (2008)
doi:10.1103/PhysRevD.77.056005
[arXiv:0707.3156 [hep-ph]].
%48 citations counted in INSPIRE as of 08 Jul 2020

%\cite{Guo:2008yz}
\bibitem{Guo:2008yz}
P.~Guo, A.~P.~Szczepaniak, G.~Galata, A.~Vassallo and E.~Santopinto,
%``Heavy quarkonium hybrids from Coulomb gauge QCD,''
Phys. Rev. D \textbf{78}, 056003 (2008)
doi:10.1103/PhysRevD.78.056003
[arXiv:0807.2721 [hep-ph]].
%94 citations counted in INSPIRE as of 23 Apr 2020


  %\cite{TorresRincon:2010fu}
\bibitem{TorresRincon:2010fu} 
  J.~M.~Torres-Rincon and F.~J.~Llanes-Estrada,
  %``Heavy Quark Fluorescence,''
  Phys.\ Rev.\ Lett.\  {\bf 105}, 022003 (2010)
  doi:10.1103/PhysRevLett.105.022003;
  %%CITATION = doi:10.1103/PhysRevLett.105.022003;%%

%\cite{Xie:2013uha}
\bibitem{Xie:2013uha}
W.~Xie, L.~Mo, P.~Wang and S.~R.~Cotanch,
%``Coulomb gauge model for hidden charm tetraquarks,''
Phys. Lett. B \textbf{725}, 148-152 (2013)
doi:10.1016/j.physletb.2013.07.003
[arXiv:1302.5737 [hep-ph]].
%8 citations counted in INSPIRE as of 15 May 2020

%\cite{Guo:2014zva}
\bibitem{Guo:2014zva}
P.~Guo, T.~Yepez-Martinez and A.~P.~Szczepaniak,
%``Charmonium meson and hybrid radiative transitions,''
Phys. Rev. D \textbf{89}, no.11, 116005 (2014)
doi:10.1103/PhysRevD.89.116005
[arXiv:1402.5863 [hep-ph]].
%13 citations counted in INSPIRE as of 15 May 2020

%\cite{Amor-Quiroz:2017jhs}
\bibitem{Amor-Quiroz:2017jhs}
D.~Amor-Quiroz, T.~Yepez-Martinez, P.~Hess, O.~Civitarese and A.~Weber,
%``Low energy meson spectrum from a QCD approach based on many-body methods,''
Int. J. Mod. Phys. E \textbf{26}, no.12, 1750082 (2017)
doi:10.1142/S0218301317500823
[arXiv:1704.01947 [nucl-th]].
%6 citations counted in INSPIRE as of 15 May 2020

\cite{Abreu:2019adi}
\bibitem{Abreu:2019adi} 
  L.~M.~Abreu, A.~G.~Favero, F.~J.~Llanes-Estrada and A.~G.~S\'anchez,
  %``Mixing and $m_q$ dependence of axial vector mesons in the Coulomb gauge QCD model,''
  Phys.\ Rev.\ D {\bf 100}, no. 11, 116012 (2019)
  doi:10.1103/PhysRevD.100.116012
  [arXiv:1908.11154 [hep-ph]].
  %%CITATION = doi:10.1103/PhysRevD.100.116012;%%
 

%\cite{Abreu:2020wio}
\bibitem{Abreu:2020wio}
L.~M.~Abreu, F.~M.~da Costa Júnior and A.~G.~Favero,
%``Revisiting the tensor $J^{PC} = 2^{--}$ meson spectrum,''
Phys. Rev. D \textbf{101}, no.11, 116016 (2020)
doi:10.1103/PhysRevD.101.116016
[arXiv:2004.10736 [hep-ph]].
%0 citations counted in INSPIRE as of 08 Jul 2020







%\cite{Blundell:1995au}
\bibitem{Blundell:1995au} 
  H.~G.~Blundell, S.~Godfrey and B.~Phelps,
  %``Properties of the strange axial mesons in the relativized quark model,''
  Phys.\ Rev.\ D {\bf 53}, 3712 (1996)
  doi:10.1103/PhysRevD.53.3712
  [hep-ph/9510245].
  %%CITATION = doi:10.1103/PhysRevD.53.3712;%%
  %57 citations counted in INSPIRE as of 08 Jun 2019

%\cite{Lytle:2018evc}
\bibitem{Lytle:2018evc}
A.~T.~Lytle, C.~T.~H.~ Davies, D.~Hatton, G.~P.~Lepage, C.~Sturm,
%``Determination of quark masses from $\mathbf{n_f=4}$ lattice QCD and the RI-SMOM intermediate scheme,''
Phys. Rev. D \textbf{98}, no.1, 014513 (2018)
doi:10.1103/PhysRevD.98.014513
[arXiv:1805.06225 [hep-lat]].
%14 citations counted in INSPIRE as of 06 May 2020

%\cite{Hatton:2020qhk}
\bibitem{Hatton:2020qhk}
D.~Hatton, C.~Davies, B.~Galloway, J.~Koponen, G.~Lepage and A.~Lytle,
%``Charmonium properties from lattice QCD + QED: hyperfine splitting, $J/\psi$ leptonic width, charm quark mass and $a_{\mu}^c$,''
[arXiv:2005.01845 [hep-lat]].
%0 citations counted in INSPIRE as of 06 May 2020

 
%\cite{Wei:2010zza}
\bibitem{Wei:2010zza} 
  K.~W.~Wei and X.~H.~Guo,
  %``Mass spectra of doubly heavy mesons in Regge phenomenology,''
  Phys.\ Rev.\ D {\bf 81}, 076005 (2010).
  doi:10.1103/PhysRevD.81.076005
  %%CITATION = doi:10.1103/PhysRevD.81.076005;%%
  %16 citations counted in INSPIRE as of 08 Apr 2020
  
  %\cite{Ebert:2011jc}
\bibitem{Ebert:2011jc} 
  D.~Ebert, R.~N.~Faustov and V.~O.~Galkin,
  %``Spectroscopy and Regge trajectories of heavy quarkonia and $B_c$ mesons,''
  Eur.\ Phys.\ J.\ C {\bf 71}, 1825 (2011)
  doi:10.1140/epjc/s10052-011-1825-9
  [arXiv:1111.0454 [hep-ph]].
  %%CITATION = doi:10.1140/epjc/s10052-011-1825-9;%%
  %100 citations counted in INSPIRE as of 08 Apr 2020

%\cite{Chen:2018hnx}
\bibitem{Chen:2018hnx}
  J.~K.~Chen,
  %``Regge trajectories for heavy quarkonia from the quadratic form of the spinless Salpeter-type equation,''
  Eur.\ Phys.\ J.\ C {\bf 78} (2018) no.3,  235.
  doi:10.1140/epjc/s10052-018-5718-z
  %%CITATION = doi:10.1140/epjc/s10052-018-5718-z;%%
  %2 citations counted in INSPIRE as of 08 Apr 2020

%\cite{Chen:2018bbr}
\bibitem{Chen:2018bbr} 
  J.~K.~Chen,
  %``Concavity of the meson Regge trajectories,''
  Phys.\ Lett.\ B {\bf 786}, 477 (2018)
  doi:10.1016/j.physletb.2018.10.022
  [arXiv:1807.11003 [hep-ph]].
  %%CITATION = doi:10.1016/j.physletb.2018.10.022;%%
  %6 citations counted in INSPIRE as of 08 Apr 2020

%\cite{Jia:2018vwl}
\bibitem{Jia:2018vwl} 
  D.~Jia and W.~C.~Dong,
  %``Regge-like spectra of excited singly heavy mesons,''
  Eur.\ Phys.\ J.\ Plus {\bf 134}, no. 3, 123 (2019)
  doi:10.1140/epjp/i2019-12474-8
  [arXiv:1811.04214 [hep-ph]].
  %%CITATION = doi:10.1140/epjp/i2019-12474-8;%%
  %4 citations counted in INSPIRE as of 08 Apr 2020


%\cite{Anisovich:2000kxa}
\bibitem{Anisovich:2000kxa} 
  A.~V.~Anisovich, V.~V.~Anisovich and A.~V.~Sarantsev,
  %``Systematics of q anti-q states in the (n, M**2) and (J, M**2) planes,''
  Phys.\ Rev.\ D {\bf 62}, 051502(R) (2000)
  doi:10.1103/PhysRevD.62.051502
  [hep-ph/0003113].
  %%CITATION = doi:10.1103/PhysRevD.62.051502;%%
  %271 citations counted in INSPIRE as of 07 Apr 2020
  
  
%\cite{Anisovich:2000ut}
\bibitem{Anisovich:2000ut} 
  A.~V.~Anisovich {\it et al.},
  %``I = 0 C = +1 mesons from 1920 to 2410 MeV,''
  Phys.\ Lett.\ B {\bf 491}, 47 (2000)
  doi:10.1016/S0370-2693(00)01018-2
  [arXiv:1109.0883 [hep-ex]].
  %%CITATION = doi:10.1016/S0370-2693(00)01018-2;%%
  %79 citations counted in INSPIRE as of 08 Apr 2020
 
 
%\cite{Afonin:2007aa}
\bibitem{Afonin:2007aa} 
  S.~S.~Afonin,
  %``Properties of new unflavored mesons below 2.4-GeV,''
  Phys.\ Rev.\ C {\bf 76}, 015202 (2007)
  doi:10.1103/PhysRevC.76.015202
  [arXiv:0707.0824 [hep-ph]].
  %%CITATION = doi:10.1103/PhysRevC.76.015202;%%
  %57 citations counted in INSPIRE as of 08 Apr 2020


%\cite{Peset:2018jkf}
\bibitem{Peset:2018jkf}
C.~Peset, A.~Pineda and J.~Segovia,
%``P-wave heavy quarkonium spectrum with next-to-next-to-next-to-leading logarithmic accuracy,''
Phys. Rev. D \textbf{98}, no.9, 094003 (2018)
doi:10.1103/PhysRevD.98.094003
[arXiv:1809.09124 [hep-ph]].
%10 citations counted in INSPIRE as of 18 May 2020


             
 
             
   
\end{thebibliography}
\end{document}